\documentclass[aps,prd,nofootinbib,superscriptaddress,floatfix]{revtex4-2}
\pdfoutput=1

\usepackage{amssymb}
\usepackage{amsmath}
\usepackage{graphicx}
\usepackage{verbatim}
\usepackage{amsfonts}
\usepackage{braket}
\usepackage{hyperref}
\usepackage{color}
\usepackage[dvipsnames]{xcolor}
\usepackage{comment}
\usepackage{graphics}
\usepackage{subfigure}
\usepackage{multirow}
\usepackage{cancel}
\usepackage{booktabs}

\allowdisplaybreaks

\newcommand{\BDD}[0]{$\bar B\to \bar D D$}

\newcommand{\SU}[0]{SU(3)$_F$}

\definecolor{darkgreen}{rgb}{0.0,0.6,0.0}

\makeatletter
\g@addto@macro\bfseries{\boldmath}
\let\Hy@backout\@gobble
\makeatother

\begin{document}

\title{$\bar{B}\rightarrow \bar{D} D$ Decays and the Extraction of $f_d/f_u$ at Hadron Colliders}

\author{Jonathan Davies}
\email{jonathan.davies-7@manchester.ac.uk}
\affiliation{Department of Physics and Astronomy, University of Manchester, Manchester M13 9PL, United Kingdom}

\author{Martin Jung}
\email{martin.jung@unito.it}
\affiliation{
Dipartimento di Fisica, Universit\`a di Torino \& INFN, Sezione di Torino, I-10125 Torino, Italy }

\author{Stefan Schacht}
\email{stefan.schacht@manchester.ac.uk}
\altaffiliation[Now at: ]{Institute for Particle Physics Phenomenology, Department of Physics, Durham University, Durham DH1 3LE, UK}
\affiliation{Department of Physics and Astronomy, University of Manchester, Manchester M13 9PL, United Kingdom}

\begin{abstract}
We perform a detailed model-independent phenomenological analysis of $\bar B\to \bar D D$ decays. Employing an SU(3)$_F$ analysis including symmetry-breaking contributions together with a conservative power counting for various suppression effects, we obtain updated Standard Model predictions for all branching fractions and CP asymmetries from a fit to the available experimental data, testable at Belle~II and the LHC experiments. These results include all relevant suppressed contributions, thereby providing upper limits on subleading Standard Model (SM) effects like ``penguin pollution'', enabling searches for physics beyond the SM. Importantly, allowing in the same fit for the production fractions of charged and neutral $B$ mesons to be different, we find $f_d/f_u = 0.86\pm0.05$, which is $2.5\sigma$ away from unity, which, if confirmed, would have important phenomenological consequences.
\end{abstract}

\maketitle


\section{Introduction}

Non-leptonic $B$ decays play an essential role in understanding CP violation and the flavour sector of the Standard Model (SM) and in searches for physics beyond it (BSM). Specifically, all angles of the unitarity triangle are measured in non-leptonic $B$ decays, the measurement of which led to the confirmation of the Kobayashi-Maskawa mechanism~\cite{Kobayashi:1973fv} as the main mechanism of CP violation in the SM (see Refs.~\cite{CKM21,UTfit:2022hsi} and references therein), and ultimately the Nobel prize for Kobayashi and Maskawa. 

$b\to c\bar cs$ decays in particular allow for precision measurements of the $B_{(s)}$ mixing phases, thanks to a strong hierarchy implied by the Cabbibo-Kobayashi-Maskawa (CKM) structure of these modes, together with the ``penguin'' suppression of the subleading hadronic matrix elements \cite{Bigi:1981qs}. This structure, most famously present in the ``golden modes'' $B\to J/\psi K$ and $B_s\to J/\psi \phi$, is present in \BDD{} decays as well, leading to the same level of theoretical control \cite{Fleischer:1999pa,Xing:1999yx,Fleischer:2007zn,Gronau:2008ed, Li:2009xf,Jung:2014jfa, Bel:2015wha}.
Importantly, this allows for sensitive searches for physics beyond the SM: if no additional sources of CP violation are present in the decay amplitudes, the time-dependent CP asymmetries allow one to access the $B_s$ mixing phase including possible BSM contributions. If instead such sources are present, they are themselves BSM signals. The possibility of such signals has recently been discussed in the context of anomalies seen in related non-leptonic $B$ decays~\cite{Bordone:2020gao, Iguro:2020ndk, Cai:2021mlt, Bordone:2021cca, Endo:2021ifc,Fleischer:2021cct,Gershon:2021pnc, Lenz:2022pgw, Bhattacharya:2022akr, Amhis:2022hpm, Piscopo:2023opf, Biswas:2023pyw, Lenz:2019lvd, Jager:2017gal, Brod:2014bfa}.

For the strategies in $b\to c\bar c s$ transitions to work, it is essential to control subleading CP-violating contributions in the SM which could mimic such BSM signals, usually dubbed ``penguin pollution''. This poses a challenge, since it requires a quantitative analysis of these subleading contributions, in contrast to just relying on the generic CKM structure. A precise calculation of the relevant hadronic matrix elements from first principles remains unfeasible for the moment, specifically since these modes do not factorise in the heavy-quark limit\footnote{Formally, factorisation can be established in the case of $B\to J/\psi M$ decays, but the corrections are $\mathcal O(1)$ \cite{Beneke:2000ry}.}\cite{Beneke:1999br,Beneke:2000ry} (see, however, Refs.~\cite{Frings:2015eva,Li:2009xf}). Therefore, generally a data-driven method is used to determine the subleading amplitudes, using the \SU{} flavour symmetry of QCD or its $U$-spin subgroup \cite{Fleischer:1999pa,Ciuchini:2005mg,Faller:2008zc,Faller:2008gt,Jung:2009pb,DeBruyn:2010hh,Lenz:2011zz,Jung:2014jfa,Jung:2015yma,Bel:2015wha,Ligeti:2015yma}, which allow the use of symmetry-related $b\to c\bar cd$ transitions to access the subleading contributions, since these exhibit a much weaker CKM hierarchy, and hence a larger sensitivity. For a reliable description of the resulting set of decay modes, however, \SU-breaking corrections need to be considered~\cite{Savage:1991wu}, since they can be sizeable and have been shown in Ref.~\cite{Jung:2012mp} to bias the extraction of the penguin pollution if not handled carefully. These considerations resulted in the framework of Refs.~\cite{Jung:2012mp,Jung:2014jfa}, applying \SU{} symmetry to the full set of decays, including subleading penguin and annihilation contributions, and importantly also the relevant \SU-breaking corrections.
In addition to subleading contributions to CP asymmetries, the structure of the resulting framework has also important implications for branching fractions. Specifically, \BDD{} decays exhibit several quasi-isospin relations \cite{Jung:2014jfa} (see also Ref.~\cite{Gronau:1995hm} for earlier partial results) that again allow for precision tests of the SM.

In the present work we adapt and update the framework developed in Ref.~\cite{Jung:2014jfa}. In particular, since its publication new experimental results appeared, namely more precise measurements of the CP asymmetries $A_{CP}(B^-\rightarrow D^- D^0)$~\cite{LHCb:2018uli, LHCb:2023wbb}, $A_{CP}(\overline{B}^0\rightarrow D^-D^+)$~\cite{LHCb:2016inx}, $S_{CP}(\overline{B}^0\rightarrow D^-D^+)$~\cite{LHCb:2016inx}, as well as measurements of $A_{CP}(B^-\rightarrow D_s^- D^0)$~\cite{LHCb:2018uli, LHCb:2023wbb}. This demands a reappraisal of the global fit and an update of the theory predictions for the as-yet unmeasured CP asymmetries and branching fraction, quantifying the available parameter space for BSM physics. Furthermore, we include in the fit the ratio of production fractions for charged and neutral $B$ mesons, $f_d/f_u$, thereby improving the fit significantly. This observation challenges the common assumption of $f_d/f_u=1$, with potentially important consequences for the phenomenology of branching fraction measurements at LHCb, for instance $\mathcal B(B_s\to \mu^+\mu^-)$.

The outline of this article is as follows. In Sec.~\ref{sec:exp} we perform an extensive analysis of the available experimental data and present averages for all measured \BDD{} branching fractions and their correlations, which we obtain from a global fit, independent of any theory input. In Sec.~\ref{sec:power-counting} we review the \SU{} parametrisation including breaking effects and specify the power counting for the different contributing topological amplitudes. Sec.~\ref{sec:numerical-results} contains the phenomenological analysis, with a detailed validation of the theory assumptions applied in our framework, discussion of key observables in \BDD, and the numerical results for branching fractions and CP asymmetries. We conclude in Sec.~\ref{sec:conclusions}. Additional details are presented in the appendix.

\section{Review of Experimental Data \label{sec:exp} }

\begin{table}[ht]
    \centering
    \begin{tabular}{ccccc}
    \hline \hline
        & Observable & Fit result & Fit result ($f_d/f_u=1$)
        & PDG~\cite{PDG2022} \\
        \hline 
         1 & $\mathcal{B}(B^- \to D^- D^0)$  & $0.383 \pm 0.034$ & $0.385 \pm 0.028$  & $0.38\pm0.04$\\
         2 & $\mathcal{B}(B^- \to D_s^- D^0)$ & $9.2 \pm 0.8$ & $9.3 \pm 0.7$ & $9.0\pm0.9$\\
         3 & $\mathcal{B}(\bar{B}^0 \to D_s^- D^+)$  & $7.6 \pm 0.7$ & $7.5 \pm 0.6$ & $7.2\pm 0.8$\\
         4 & $\mathcal{B}(\bar{B}_s \to D^- D_s^+)$  & $0.280 \pm 0.044$ & $0.278 \pm 0.040$ & $0.28\pm0.05$\\
         5 & $\mathcal{B}(\bar{B}^0 \to D^- D^+)$  & $0.231 \pm 0.023$ & $0.231\pm0.023$ & $0.211\pm 0.018^\dagger$\\
         6 & $\mathcal{B}(\bar{B}_s \to D_s^- D_s^+)$  & $4.3 \pm 0.6$ & $4.3 \pm 0.6$  & $4.4\pm0.5$\\
         7 & \color{gray}{$\mathcal{B}(\bar{B}^0 \to D_s^- D_s^+)$}  &    &  & \color{gray}{$\leq 0.036$}\, \cite{Zupanc:2007pu} \\
         8 & $\mathcal{B}(\bar{B}_s \to D^- D^+)$  & $0.24 \pm 0.05$ & $0.24 \pm 0.05$ & $0.22\pm0.06$\\
         9 & $\mathcal{B}(\bar{B}^0 \to \bar{D}^0 D^0)$  & $0.012 \pm 0.006$ & $0.012\pm 0.006$  & $0.014\pm0.007$\\
         10 & $\mathcal{B}(\bar{B}_s \to \bar{D}^0 D^0)$  & $0.166 \pm 0.039$ & $0.165 \pm 0.036$  & $0.19\pm 0.05$\\  \hline \hline
    \end{tabular}
    \centering
    \includegraphics[width=0.7\textwidth]{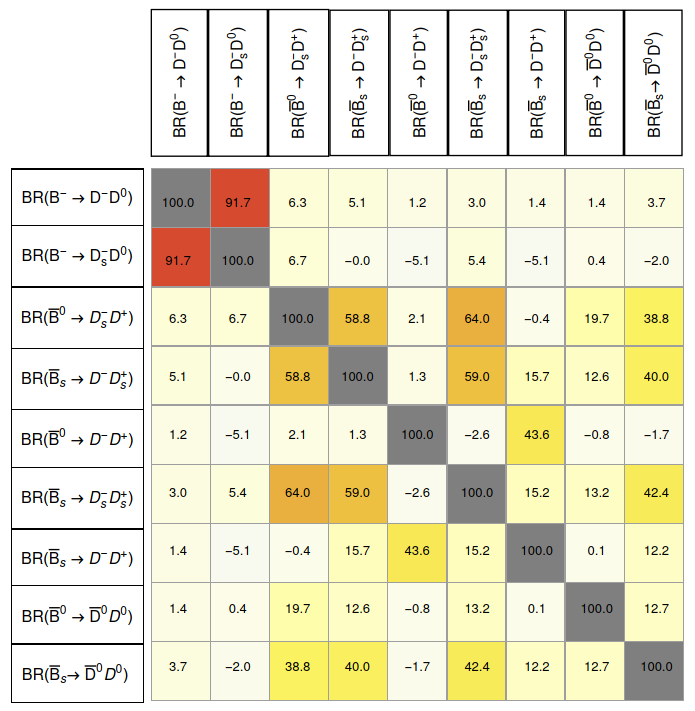}
    \caption{
    Summary of the available experimental information on \BDD{} branching fractions, given in units of $10^{-3}$, with their correlations given in percent. Our main fit results include additional uncertainties due to the production fractions (see text). The second column of results illustrates the impact of the assumption $f_d/f_u=1$, while the last column provides the present PDG values for these branching fractions for comparison. We provide the symmetric uncertainties as appropriate for the use with the correlation matrix below. This is an excellent approximation, as can be seen from the comparison with the values given in Table~\ref{tab:BRs-theory}. The limit for $\mathcal{B}(\bar{B}^0\rightarrow D_s^- D_s^+)$ is not included in the fits.
${}^\dagger$: The reduced uncertainty compared to our fit is due to the inclusion of the superseded measurement in Ref.~\cite{Belle:2007ebz}.}
    \label{tab:BRsJD}
\end{table}

We start by considering the available experimental information on the branching ratios and CP asymmetries in detail. We emphasise that at this stage no theoretical assumptions are made.

\subsection{Branching fractions \label{sec:branching-fractions}}

Instead of using the values for the branching fractions provided in Ref.~\cite{PDG2022}, we extract them ourselves in a fit to the available measurements. The reasons for this procedure are as follows:
\begin{enumerate}
\item The main point is to include significant correlations, stemming from two different sources. Firstly, correlations are introduced
by the simple fact that often ratios of branching fractions are measured instead of individual ones, especially but not
exclusively at hadron colliders. Secondly, important correlations arise due to external inputs, especially the $D$-meson branching fractions and the $B$ production fractions.
The latter often dominate the overall systematic uncertainty, specifically in $B_s$ decays, and thereby in some cases even the total uncertainty. We will see below
that some of the correlations between the different branching fractions are above $60\%$, hence their inclusion is mandatory for
a meaningful interpretation of the available data. 
\item Having made explicit the external inputs, we update them where possible (and necessary), especially in the cases
of the $D$-meson branching fractions and the different production fractions in older measurements.
\item We use a modified treatment of the different production fractions:
\begin{itemize}
    \item $f_\pm,f_{00}$: The branching fraction of the $\Upsilon(4S)$ resonance into charged and neutral $B$-meson pairs determine the numbers of the corresponding $B$-meson species at $B$-factory experiments, thereby enabling determinations of absolute branching fractions in $B$-meson decays. While na\"{i}vely the two decay rates are equal and correspond together to the total $\Upsilon(4S)$ decay rate, isospin breaking in the decay is enhanced by the fact that the $\Upsilon(4S)$ is so close to the $B\bar B$ threshold \cite{Atwood:1989em,Lepage:1990hm}. Already the phase-space difference results in a $\sim 5\%$ correction to the na\"{i}ve assumption of $f_\pm/f_{00}=1$. Since theoretical estimates for this breaking vary widely, we use the available experimental information to restrict it \cite{Jung:2012mp,Bernlochner:2023bad}. We use $f_{00}=0.484\pm0.007$, which covers the interval determined in Ref.~\cite{Bernlochner:2023bad}. Given the large uncertainties in the $B^-\to \bar D D$ branching fractions, we use the simplifying assumption $f_\pm+f_{00}=1$, since the additional uncertainty related to this assumption is negligible in comparison, even though it will be important to include it once higher precision is achieved \cite{Amhis:2022hpm,Bernlochner:2023bad}.
    \item $f_s/f_d$: Another important input is the production fraction of $B_s$ mesons relative to that of $B_d$ mesons. We use its value as extracted in semileptonic decays \cite{Bordone:2020gao, LHCb:2011leg, Storaci:2013jqy}, since the hadronic decays that are also commonly used in its extraction \cite{Fleischer:2010ca} exhibit an anomaly \cite{Bordone:2020gao}. The ratios $f_s/f_d$ and $f_{\Lambda_b}/{f_d}$ exhibit a significant dependence on the transverse momentum \cite{LHCb:2021qbv,LHCb:2019fns,CMS:2022wkk} so we include $(f_s/f_d)_{\mathrm{Tev}}$ as an independent quantity. 
    The measured ratio of branching fractions~\cite{CDF:2012xmd} then serves as an independent determination of this ratio of production fractions.     
    \item $f_d/f_u$: The relative production fraction of neutral to charged $B$ mesons at hadron colliders is commonly set to unity in experimental (and theoretical) analyses. However, the non-trivial isospin structure of both initial and final states in this case render this assumption hard to justify~\cite{Bernlochner:2023bad}. In fact, the expected value for this ratio depends on the production mechanism of the $B$ mesons, and presumably also on the transverse momentum, as observed for $f_s/f_d$ and $f_{\Lambda_b}/{f_d}$. 
    There is no measurement available yet for this ratio of production fractions at LHCb. We want to determine its size from the available \BDD{} data instead of making assumptions, and therefore allow in our fit conservatively for 
    \begin{align}
        f_d/f_u\in[0.5,1.5]\,.
    \end{align}
    This results in an additional systematic uncertainty for the related branching fractions. 
\end{itemize}
\end{enumerate}
We do not take CP asymmetries into account in this fit, since they are very weakly correlated with the branching fractions, with unknown correlations.
In the appendix we give a detailed list of the experimental results that we use in our analysis, going systematically through the relevant inputs from the different experiments~\cite{LHCb:2023wbb,LHCb:2013sad, LHCb:2014scu, CLEO:1995psi, BaBar:2006jvx, BaBar:2006uih, Belle:2008doh, Zupanc:2007pu, Belle:2012mef, Belle:2012tsw, CDF:2012xmd}.

Our fit results for the branching ratios and their correlations are given in Table~\ref{tab:BRsJD}. The fit is excellent and we do not observe any indication of significant incompatibilities between different experimental results. The correlation matrix corresponds to a Gaussian approximation for the uncertainties, which we checked explicitly to be appropriate in this case. We illustrate the impact of removing the assumption of $f_d/f_u=1$ by showing also the results with this assumption in place, and we compare our results to those obtained by the PDG~\cite{PDG2022}. Regarding the latter comparison, in principle we expect larger uncertainties, due to the additional uncertainties from varying the production fractions $f_d/f_u$ and $f_{\pm,00}$. However, updating lifetimes and $D$-meson branching fractions, as well as including a very recent LHCb analysis~\cite{LHCb:2023wbb}, results in reduced uncertainties for most branching fractions. Nevertheless, the main difference to the PDG fit is that we provide the full correlation matrix, which is crucial for a meaningful interpretation of the available data.

Regarding the production fractions, the values for $(f_s/f_d)_{\mathrm{LHCb}}$ and $f_{\pm,00}$ change only marginally, meaning that there is no significant independent information on these ratios from \BDD{} at the moment. On the other hand, we obtain 
\begin{align}
(f_s/f_d)_{\mathrm{Tev}}=0.34^{+0.06}_{-0.05}\,,
\end{align}
corresponding to a competitive independent determination of this quantity, in line with the average presented by HFLAV in Ref.~\cite{HFLAV:2019otj}, and about one standard deviation higher than the value obtained by LHCb. Furthermore, we obtain
\begin{align}
(f_d/f_u)^{\text{LHCb,~7 TeV}} &= 0.99^{+0.15}_{-0.13}%
\,, \label{eq:fd-over-fu-exp}
\end{align}
which agrees with the na\"{i}ve value of $f_d/f_u=1$, but also still allows for large deviations. Note that while the result in Eq.~\eqref{eq:fd-over-fu-exp} is completely model-independent, it uses a large amount of information external to LHCb. It corresponds to the first extraction of this quantity at LHCb to our knowledge (see Ref.~\cite{CMS:2022wkk} for an analysis at CMS). Below we show how to improve on this extraction by adding minimal theoretical information.
The sensitivity to $f_d/f_u$ stems mainly from the LHCb measurement of the ratio~\cite{LHCb:2013sad}
\begin{align}
\frac{f_u}{f_d} \frac{\mathcal{B}(B^-\rightarrow D^0 D_s^-)}{\mathcal{B}(\bar{B}^0 \rightarrow D^+ D_s^-)}\,, \label{eq:fufd-sensitivity}
\end{align}
since the individual branching fractions in this ratio are also being determined at the $B$ factories.
Ref.~\cite{LHCb:2013sad} also provides the ratios
\begin{align}
\frac{f_s}{f_d} \frac{f_d}{f_u}
\frac{\mathcal{B}(\bar{B}_s\rightarrow D^0 \bar{D}^0)}{\mathcal{B}(B^-\rightarrow D^0 D_s^-)}\, \quad \mathrm{and} \quad 
\frac{f_d}{f_u} \frac{\mathcal{B}(\bar{B}^0\rightarrow D^0 \bar{D}^0)}{\mathcal{B}(B^-\rightarrow D^0 D_s^-)}\,.
\end{align}
Since the numerators of these ratios have not been determined individually, the corresponding branching fractions will be affected by changes in $f_d/f_u$.

\subsection{CP asymmetries}

For the CP asymmetries we use the notation of Ref.~\cite{Jung:2014jfa}, which we briefly summarise as follows.
We write a time-dependent CP asymmetry as 
\begin{align}
a_{CP}(\mathcal{D}; t) &= \frac{
    \Gamma(\mathcal{D}; t) - \Gamma(\overline{\mathcal{D}}, t)
    }{
    \Gamma(\mathcal{D}; t) + \Gamma(\overline{\mathcal{D}}, t)
    }=\frac{
    S_{CP}(\mathcal{D})\sin(\Delta m t) + A_{CP}(\mathcal{D}) \cos(\Delta m t)
    }{
    \cosh(\Delta \Gamma t/2) - A_{\Delta \Gamma}(\mathcal{D}) \sinh(\Delta \Gamma t/2)
    }\,,
\end{align}
where $\mathcal D$ denotes a $\bar B$-meson decay (containing a $b$ quark)\footnote{Therefore, our definition of $\lambda$ agrees with the PDG one \cite{PDG2022}, as we have
$\mathcal A(\mathcal D)=(\bar A_f)_{\mathrm{PDG}}$.}, $\Delta \Gamma$ the decay-width difference of the decaying $B$ meson, and with
\begin{align}
A_{CP}(\mathcal{D}) &= -\frac{1 - \vert \lambda(\mathcal{D})\vert^2}{1 + \vert \lambda(\mathcal{D})\vert^2}\,, &
S_{CP}(\mathcal{D}) &= \frac{2\, \mathrm{Im} \lambda(\mathcal{D})}{1 + \vert \lambda(\mathcal{D})\vert^2}\,, &
A_{\Delta\Gamma}(\mathcal{D}) &=  \frac{2\, \mathrm{Re} \lambda(\mathcal{D})}{1 + \vert \lambda(\mathcal{D})\vert^2}\,, \quad\mbox{and}&\lambda(\mathcal{D}) &= e^{-i\phi_D} \frac{\mathcal{A}(\mathcal{D})}{\overline{\mathcal{A}}(\mathcal{D})}\,.
\end{align}
The CP eigenvalue for the decays we consider is $\eta_{CP}^f = +1$.
The weak mixing phase is given as 
\begin{align}
e^{-i\phi_D} &= \frac{V^*_{tb}V_{tD}}{V_{tb}V^*_{tD}}\,,\quad
D = d\,, \, s\,.
\end{align}
We will also use
\begin{align}
\Delta S= -\eta_{CP}^f S_{CP} - \sin\phi_D\,,
\end{align}
since this quantity has an identical power counting as $A_{CP}$ for the modes with a time-dependent CP asymmetry, unlike $S_{CP}$.

\begin{table}[t]
    \centering
    \begin{tabular}{c|c|l|l}\hline\hline
    Observable & Value/\% & Comment & Ref.\\
    \hline
$A_{CP}(B^- \to D^- D^0)$ & $-3.3 \pm 7.1$ & Average excluding LHCb result (see text).  & \cite{Belle:2008doh, BaBar:2006uih}\\
$A_{CP}(B^- \to D^- D^0)$ & $2.5 \pm 1.1$ & Only LHCb. & \multirow{2}{*}{\cite{LHCb:2023wbb}}\\
$A_{CP}(B^- \to D_s^- D^0)$& $0.5 \pm 0.6$ & $\mbox{corr}(D^-D^0,D_s^-D^0)=38.6\%$  & \\
$S_{CP}(\bar{B}^0 \to D^- D^+)$ & $-81 \pm 20$  & Our average (see text for details). & \multirow{2}{*}{\cite{LHCb:2016inx,Belle:2012mef,BaBar:2008xnt,PDG2022}}\\ 
$A_{CP}(\bar{B}^0 \to D^- D^+)$ & $13 \pm 17$ & $\mathrm{corr}(S_{CP},A_{CP})= -19.6\%$ & 
\\ 
$S_{CP}(\bar{B}_s \to D_s^- D_s^+)$ & $-2 \pm 17$ & Extracted from $|\lambda(\bar{B}_s \to D_s^- D_s^+)|$ and $\phi_s^{\mathrm{eff}}(\bar{B}_s \to D_s^- D_s^+)$. & \multirow{2}{*}{\cite{LHCb:2014ini}} \\
$A_{CP}(\bar{B}_s \to D_s^- D_s^+)$ & $-7.8 \pm 17.9$ & $\mathrm{corr}(S_{CP},A_{CP})= -3.2\%$ &
\\\hline\hline
    \end{tabular}
    \caption{Experimental input data for CP asymmetries. We do not use the value $A_{CP}(\overline{B}^0\rightarrow D_s^-D^+) = -0.01\pm 0.02$ provided in Ref.~\cite{Belle:2012mef}, as the quoted uncertainty is statistical only and the measurement therein was only performed as a cross check. 
    }
    \label{tab:CPasym-exp}
\end{table}

\begin{figure}[t]
\includegraphics[width=7cm]{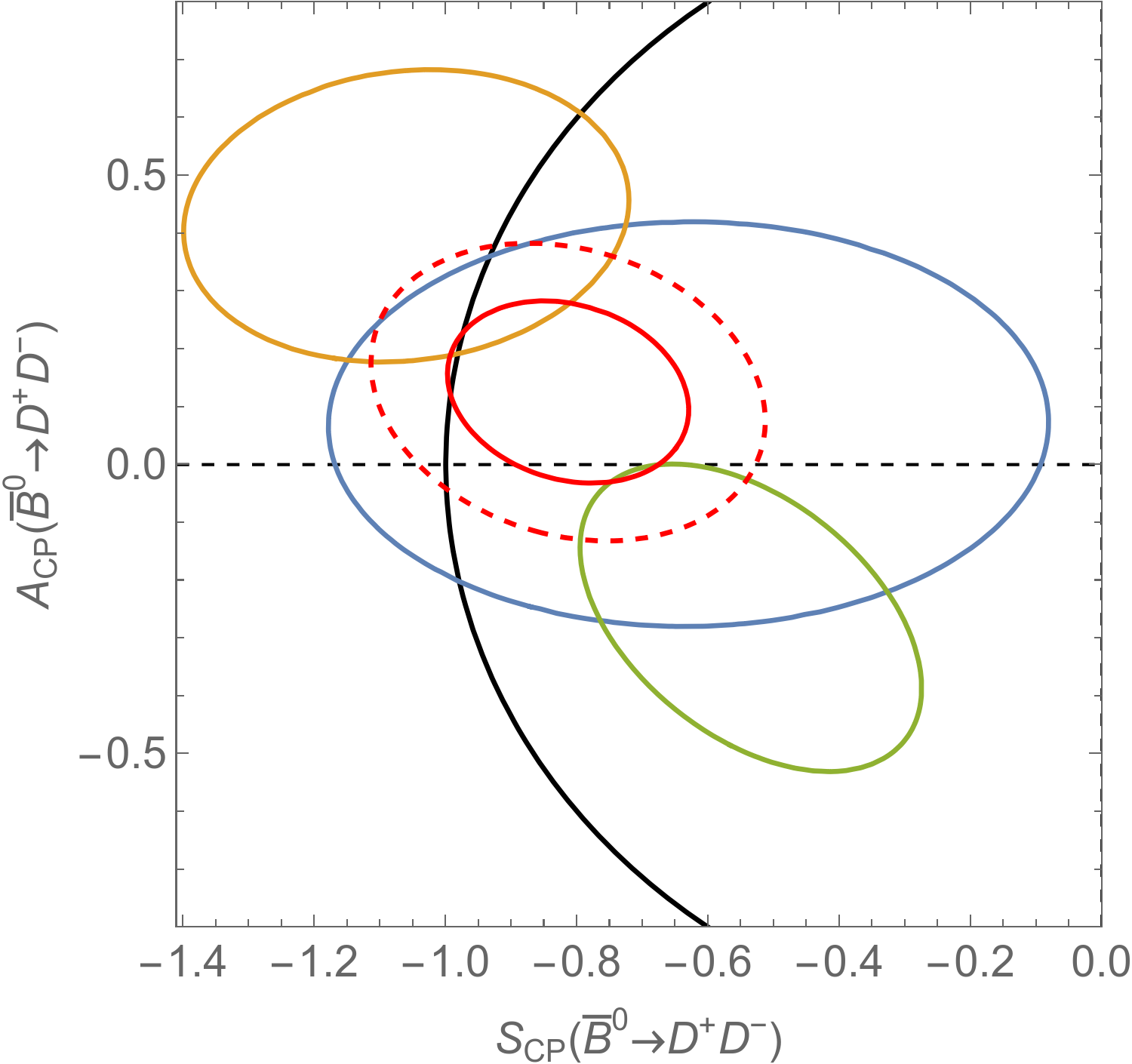}
\caption{\label{fig:ACPB0D+D-}Time-dependent CP asymmetry measurements of $\bar{B}^0\to D^-D^+$ from 
Belle~\cite{Belle:2012mef} (yellow), 
BaBar~\cite{BaBar:2008xnt} (blue) and
LHCb~\cite{LHCb:2016inx} (green), together with their standard correlated average (red solid). Our average with the uncertainties
symmetrically enlarged, as described in the text, is also shown (red dashed) and in our analysis we employ this average. All data given correspond to the $68\%$~CL regions. The black solid line marks the limit of the physical parameter space which is defined as $A_{CP}(\bar{B}^0\to D^-D^+)^2 + S_{CP}(\bar{B}^0\to D^-D^+)^2 \leq 1$.}
\end{figure}

We summarise the employed measurements of CP asymmetries in Table~\ref{tab:CPasym-exp}, including the available correlations.
The CP asymmetries in charged $B$-meson decays are treated as follows: we average the measurements included in the PDG average \cite{PDG2022}, but excluding the superseded LHCb measurements in Ref.~\cite{LHCb:2018uli}. Then we add the correlated LHCb measurements from Ref.~\cite{LHCb:2023wbb}, thereby accounting for the corresponding correlation and avoiding double-counting of LHCb inputs.
For the measurements of $A_{CP}(\bar{B}^0\rightarrow D^-D^+)$ and $S_{CP}(\bar{B}^0\rightarrow D^-D^+)$, we observe a tension between the measurements by Belle~\cite{Belle:2012mef} and BaBar~\cite{BaBar:2008xnt}. We increase the uncertainty of the resulting average, according to a trivial generalisation of the so-called PDG prescription: we perform a weighted average of the three available measurements, including their correlations~\cite{Schmelling:1994pz}.
We then apply the same scaling factor $S = \sqrt{\chi^2/N_{\mathrm{dof}}}\approx 1.6$ to each of the three pairs of uncertainties, thereby preserving the overall correlation between them and the final result. This procedure is visualised in Fig.~\ref{fig:ACPB0D+D-} and results in the uncertainties given in Table~\ref{tab:CPasym-exp}.

\section{\SU{} Expansion and Power Counting \label{sec:power-counting}}

\begin{table}[t]
    \centering
    \begin{tabular}{cc|cccccc|cccc}\hline\hline
    &&\multicolumn{6}{|c|}{\SU{} Limit} & \multicolumn{4}{c}{\SU{} Breaking}\\
    \hline \hline
        & Mode\textbackslash Amplitude &  $T$ & $A^c$ & $\tilde{P_1}$ & $\tilde{P_3}$ & $A_1^u$ & $A_2^u$ & $\delta T_1$ & $\delta T_2$ & $\delta A^c_1$ & $\delta A^c_2$ \\ \hline
        & Counting &  1 & $\epsilon^{1.5}$ & $\epsilon^{2.5}$ & $\epsilon^{3.5}$ & $\epsilon^{2.5}$ & $\epsilon^{3.5}$ &  $\epsilon^{1(2)}$ & $\epsilon^{1(2)}$ & $\epsilon^{2.5}$ & $\epsilon^{2.5}$ \\ \hline
        1& $B^- \to D^- D^0$ & 1 & 0 & $-1$ & 0 & 1 & 0 & 0 & $-\frac{1}{2}$ & 0 & 0\\
        2& $B^- \to D_s^- D^0$ & 1 & 0 & $-1$ & 0 & 1 & 0 & 1 & 0 & 0 & 0 \\
        3& $\bar{B}^0 \to D_s^- D^+$ & 1 & 0 & $-1$ & 0 & 0 & 0 & 1 & 0 & 0 & 0\\
        4& $\bar{B}_s \to D^- D_s^+$ & 1 & 0 & $-1$ & 0 & 0 & 0 & $-1$ & $\frac{1}{2}$ & 0 & 0 \\
        5& $\bar{B}^0 \to D^- D^+$ & 1 & 1 & $-1$ & $-1$ & 0 & 0 & 0 & $-\frac{1}{2}$ & $\frac{1}{2}$ & $-\frac{1}{2}$\\
        6& $\bar{B}_s \to D_s^- D_s^+$ & 1 & 1 & $-1$ & $-1$ & 0 & 0 & 0 & 1 & $-1$ & $1$ \\
        7& $\bar{B}^0 \to D_s^- D_s^+$ & 0 & 1 & 0 &  $-1$ & 0 & 0 & 0 & 0 & $\frac{1}{2}$ &  $\frac{1}{2}$\\
        8& $\bar{B}_s \to D^- D^+$ & 0 & 1 & 0 &  $-1$ & 0 & 0 & 0 & 0 & $-1$ &  0\\
        9& $\bar{B}^0 \to \bar{D}^0 D^0$ & 0 & $-1$ & 0 & 1 & 0 & $-1$ & 0 & 0 & $-\frac{1}{2}$ & $\frac{1}{2}$\\
        10& $\bar{B}_s \to \bar{D}^0 D^0$ & 0 & $-1$ & 0 & 1 & 0 & $-1$ & 0 & 0 & 1 & 0 \\
         \hline \hline
    \end{tabular}
    \caption{\SU{} limit decomposition (left side) and parametrisation of \SU{}-breaking contributions (right side) for the \BDD{} system and power counting for the topological amplitudes with partially absorbed CKM factors (as in Eq.~\ref{eq:topo_decomposition_CKM}). Numbers in parentheses indicate the power counting for the imaginary parts of the SU(3)$_F$-breaking tree amplitudes, which are additionally suppressed (see text for details). Table adapted from Ref.~\cite{Jung:2014jfa}. \label{tab:topo_table}}
\end{table}

\begin{table}[t]
    \centering
    \begin{tabular}{ccccccc}\hline\hline
    Topology & CKM & Colour & Annihilation & Penguin & \cancel{$SU(3)_F$} & Total \\
                \hline 
                $\vert T\vert$ & 1 &1 & 1 & 1 & 1 & 1\\
                $\vert A^c\vert$ & 1 & $\epsilon$ & $\epsilon^{0.5}$ & 1 & 1&  $\epsilon^{1.5}$\\
                $\vert \tilde{P}_1\vert$ & $\epsilon$ & $\epsilon$ & 1 & $\epsilon^{0.5}$ & 1 &$\epsilon^{2.5}$ \\
                $\vert \tilde{P}_3\vert$ & $\epsilon$ & $\epsilon^{2}$ & 1 & $\epsilon^{0.5}$ & 1 & $\epsilon^{3.5}$\\
                $\vert A^u_1\vert$ & $\epsilon$ & 1 & $\epsilon^{1.5}$ & 1 & 1& $\epsilon^{2.5}$\\
                $\vert A^u_2\vert$ & $\epsilon$ & $\epsilon$ & $\epsilon^{1.5}$ & 1 & 1 & $\epsilon^{3.5}$\\
                $\vert \delta T_{1,2}\vert$ & 1 & 1\, ($\epsilon$) & 1 & 1 & $\epsilon$ & $\epsilon \, (\epsilon^2)$\\
                 $\vert \delta A^c_{1,2}\vert$ & 1 & $\epsilon$ & $\epsilon^{0.5}$ & 1 & $\epsilon$ &  $\epsilon^{2.5}$\\
                 \hline \hline
                
    \end{tabular}
    \caption{Anatomy of suppression factors for the magnitude of topological amplitudes. The imaginary part of the \SU{}-breaking amplitudes $\delta T_{1,2}$ have additional suppression (see Eq.~(\ref{eq:power-counting-deltaT})). The total power counting is obtained by multiplying the different contributions in a given row. This means that an entry of \lq\lq{}1\rq\rq{} implies that no corresponding suppression applies.}
    \label{tab:power-counting}
\end{table}

In order to obtain theoretical expressions for the observables described in the previous section, we need the hadronic matrix elements of the corresponding effective Hamiltonian between the initial $B$ meson state and the final $\bar D D$ pair~\cite{Buchalla:1995vs},
\begin{align}
\mathcal{H}_{\mathrm{eff}}^{b\rightarrow d,s} &= 
    \frac{4 G_F}{\sqrt{2}} \sum_{U=c,u} \sum_{D=d,s} \lambda_{UD} 
        \left(\sum_{i=1}^2 C_i \mathcal{O}_i^U + \sum_{i=3}^{10} C_i \mathcal{O}_i\right)\,.\label{eq:Hamiltonian}
 \end{align}
Here $\lambda_{UD}=V_{Ub}V_{UD}^*$ denotes a product of CKM-matrix elements ($U=t$ has been eliminated using the unitarity of the CKM matrix), $\mathcal O_{1,2}^U$ are tree operators, $\mathcal O_{3-6}$ penguin operators, and $\mathcal O_{7-10}$ electroweak penguin operators. Importantly, the Wilson coefficients $C_i$ obey the hierarchy $C_1\sim C_2/N_C\gg C_{3-6}\gg C_{7-10}$. 

Presently there is no reliable way to calculate the necessary hadronic matrix elements for \BDD{} decays from first principles. Instead, we follow the previous analysis performed by two of us~\cite{Jung:2014jfa}, and employ the approximate \SU{} symmetry of QCD which \emph{relates} all ten decay modes under consideration. We include the leading \SU-breaking corrections following Ref.~\cite{Savage:1991wu} (see also Refs.~\cite{Zeppenfeld:1980ex,Chau:1982da,Savage:1989ub,Grinstein:1996us}). This is since corrections to the symmetry limit are expected to be sizeable, scaling like $m_s/\Lambda\sim 30\%$, where $\Lambda$ denotes a non-perturbative hadronic scale. We then group these \SU-amplitudes into scale- and scheme-independent \emph{topological amplitudes}~\cite{Buras:1998ra} (see Ref.~\cite{Jung:2014jfa} for a translation). 
We write the resulting decay amplitude for a decay $\mathcal D$ as (see also the point ``CKM structure'' below)
\begin{align}
    \mathcal A(\mathcal D) &= \mathcal A_c(\mathcal D) + \mathcal A_u(\mathcal D) \\
        &= \frac{\lambda_{cD}}{\lambda_{cs}} \sum_iC^i_c(\mathcal D)\mathcal{T}^i_c + \frac{\lambda_{uD}}{R_u \lambda_{cs}} \sum_i C_u^i(\mathcal D)\mathcal{T}^i_u\,,
        \label{eq:topo_decomposition_CKM}
\end{align}
where $\mathcal{T}^i_{c,u}$ are the topological amplitudes 
with partially absorbed CKM factors, for which we show the coefficients $C_{c,u}^i(\mathcal D)$ in Table~\ref{tab:topo_table}.
The topological amplitudes capture the flavour- and colour-flow of a given hadronic matrix element, but are all-order quantities in the strong coupling. This translation does not reduce the number of unknown parameters, but allows the application of power-counting arguments to the resulting amplitudes, according to their topological structure. This power counting is adapted from Ref.~\cite{Jung:2014jfa}, and modified in order to be even more conservative (see details given below). The different suppression effects are categorised in terms of powers of a generic suppression factor $\epsilon\sim 30\%$.

\begin{itemize}
\item \textbf{CKM structure:} 
We use the relations 
\begin{align}
  \lambda_{cd}=-\bar\lambda \lambda_{cs} \,,\quad 
  \lambda_{ud}\equiv -R_u e^{-i\gamma}\lambda_{cd}=R_u\bar\lambda e^{-i\gamma}\lambda_{cs} \,\quad\mbox{and}\quad
  \lambda_{us}=\bar\lambda^2R_ue^{-i\gamma}\lambda_{cs}\,,
\end{align}
which hold up to sub-percent level. 
Using these relations, $\lambda_{cs}$ becomes a common normalisation factor and can be absorbed. We also absorb the factor $R_u$ into the matrix elements proportional to $\lambda_{uD}$,  and count this as
\begin{align}
R_u\sim\epsilon\,. 
\end{align}
We include explicitly the relevant factors of $\bar\lambda=\lambda(1+\lambda^2/2)$, where $\lambda$ denotes the Wolfenstein parameter. 
\item \textbf{$\mathbf{SU(3)_F}$ structure:} We consider $SU(3)_F$-breaking contributions generically to scale as 
\begin{align}
\mathcal{O}(\epsilon)\,,  
\end{align}
with one exception described under \lq\lq{}Colour suppression\rq\rq{}.
\item \textbf{Colour suppression:} We include the relative counting in $1/N_C\sim \epsilon$ for the topological amplitudes, following Refs.~\cite{tHooft:1973alw,  Buras:1985xv, Buras:1998ra, Jung:2014jfa}. This implies 
\begin{align}
T,A_1^u\sim1\,, \qquad
\tilde P_1,A_2^u,A^c\sim1/N_C\,, \qquad 
\tilde P_3\sim1/N_C^2\,. \label{eq:colour-suppression} 
\end{align}
We note that in the
large-$N_C$ limit the tree amplitudes factorise even beyond the $SU(3)_F$ limit, with corrections to this limit scaling
like $1/N_C^2$. While we do not use the factorised amplitudes explicitly, we take this general structure into account in
a conservative manner by assuming that the real parts and absolute values of the \SU-breaking corrections to tree diagrams are suppressed as $\mathcal{O}(\epsilon)$, while the corresponding imaginary parts are suppressed as~$\mathcal{O}(\epsilon^2)$, \emph{i.e.}:
\begin{align}
\vert  \delta T_{1,2}\vert \sim \mathcal{O}(\epsilon)\,,\qquad
\mathrm{Re}(\delta T_{1,2}) \sim \mathcal{O}(\epsilon)\,, \qquad
\mathrm{Im}(\delta T_{1,2}) \sim \mathcal{O}(\epsilon^2)\,.  \label{eq:power-counting-deltaT}
\end{align}
\item \textbf{Penguin suppression:} The penguin amplitudes receive two different contributions. One is from
tree matrix elements of penguin operators, which have a quantifiable suppression from their Wilson coefficients that
would correspond to a scaling of $\mathcal{O}(\epsilon^2)$. The other contribution stems from penguin matrix elements of
tree operators, and here the suppression is much harder to quantify. While we do not see significant indications of a large
penguin amplitude in \BDD{} decays, the central values of the CP asymmetries in $\bar B\to D^+D^-$ especially are sizeable.
To be conservative, we therefore assign a suppression factor of
\begin{align}
\epsilon^{1/2}\sim55\%
\end{align}
to this effect. We deem this counting, which lies between the two scenarios ``Standard Counting'' and ``Enhanced Penguins'' in Ref.~\cite{Jung:2014jfa}, sufficiently conservative that a significant violation would be indicative of BSM physics.
Under \SU{}, the penguin operators transform simply as triplets, as does the part of the electroweak penguin operators involving charm quarks. These operators contribute to the same reduced matrix elements, included in our analysis below. The remaining contributions are suppressed not only by their tiny Wilson coefficients, but involve an additional suppression of their matrix elements, rendering them completely negligible (see also Ref.~\cite{Jung:2014jfa}). This holds also for the matrix elements of tree operators contributing to the same structure, since they necessarily involve photon interactions $\mathcal O(\alpha)\sim \mathcal O(\epsilon^4)$.  
\item \textbf{Annihilation:} The annihilation contributions, involving the spectator quark, involve na\"{i}vely a suppression of $\mathcal O(\Lambda/m_b)$, but with heavy mesons in the final state this na\"{i}ve estimate breaks down. We assign a suppression factor of 
\begin{align}
\epsilon^{1/2}\sim 55\%
\end{align}
to annihilation contributions in general, which is slightly more conservative than what we considered in Ref.~\cite{Jung:2014jfa}. We add an additional factor
$\epsilon$ in the cases where the process requires the creation of a $c\bar c$ pair from the vacuum, since such contributions vanish in the heavy-quark limit \cite{Mannel:1990un}. This applies to the annihilation topologies $A_1^u$ and $A_2^u$, which therefore receive an annihilation-suppression of~$\epsilon^{1.5}$.
\end{itemize}

We summarise the decomposition of the power counting for the different parameters in Table~\ref{tab:power-counting}, and for convenience also give the total suppression factors for each amplitude in Table~\ref{tab:topo_table}. 
The amplitudes not listed in Table~\ref{tab:power-counting} are not included in our phenomenological analysis, but are given in Ref.~\cite{Jung:2014jfa}. The reasoning for this is as follows:
\begin{itemize}
    \item The only missing matrix element in $\mathcal A_c$ is second-order in the \SU{}-breaking, and can be absorbed in nine out of ten decay modes into matrix elements already present in our fit. The only mode it contributes to independently is $\bar B_s\to D_s^+D_s^-$, that is already well-described in our fits and has already several \SU-breaking amplitudes, some of which are only restricted by our power counting. We expect therefore that the inclusion of this matrix element will only slightly increase the uncertainties for this mode, and otherwise have no effect on the fit.
    \item The \SU-breaking contributions to $\mathcal A_u$ are not included, since already the larger matrix elements at leading order in \SU{} are not restricted by existing measurements. Given the overall suppression of $\mathcal A_u$ compared to $\mathcal A_c$, the effect of the neglected matrix elements on branching fractions is negligible, while the uncertainty for the CP asymmetries would be increased by about $30\%$ by their inclusion.
\end{itemize}
In order to allow for some enhancement even beyond the conservatively chosen nominal suppression expected from the effects discussed above, we reduce the total suppression factor for each topological amplitude but the leading tree amplitude by an additional factor of $\epsilon^{-1/2}$. 

\section{Phenomenological Analysis \label{sec:numerical-results}}

We use the available experimental data, summarised in Tables~\ref{tab:BRsJD} and~\ref{tab:CPasym-exp}, to perform two tasks. The first is to validate our theory assumptions by verifying the generic predictions of our setup developed in the previous section, while the second is a global fit, providing additional consistency checks and, more importantly, quantitative predictions for all observables of all decay modes, including those not yet measured. The required numerical input from the global CKM fit~\cite{CKM21} is given in Table~\ref{tab:CKMinput}. When values are given without uncertainties, the central values are used. In the case of $y_s$ and $\phi_s$, the uncertainties are negligible compared to other experimental and/or theoretical uncertainties included in the fit. In the case of the CKM phase $\gamma$, the reason lies in the so-called \emph{reparametrisation invariance} \cite{London:1999iv,Botella:2005ks,Feldmann:2008fb,Jung:2014jfa}. This effect, discussed in the context of broken \SU{} in detail in Ref.~\cite{Jung:2014jfa}, implies the equivalence of a change in the apparent phase $\gamma$ in \BDD{} and a shift in the various topological amplitudes. This therefore renders these modes insensitive to this phase, unless quantitative theory assumptions are made about at least one of the topological amplitudes (see also the discussions in Refs.~\cite{Gronau:2008ed} and \cite{Jung:2014jfa}). Since we allow very large ranges for the amplitudes, the uncertainty for $\gamma$ becomes irrelevant and only slightly changes the interpretation of the fitted topological amplitudes. 

\begin{table}[t]
    \centering
    \begin{tabular}{c|c}\hline\hline
    Input & Value \\
    \hline
$\lambda$ & $0.22500$ \\
$\gamma$ & $1.143$ \\
$\phi_d \equiv 2 \beta$ & $0.787\pm 0.018$ \\
$\phi_s\equiv -2\beta_s$ & $-2\times 0.01841$\\
$y_s\equiv \Delta \Gamma_s/(2 \Gamma_s)$ & $0.062$ \\\hline\hline
\end{tabular}
\caption{Additional input data from the global CKM fit~\cite{CKM21}.}
\label{tab:CKMinput}
\end{table}

\subsection{Validation of Theory Assumptions \label{sec:test-assumptions}}

\begin{table}[]
    \centering
    \begin{tabular}{l|c c}\hline\hline
        Relative rate & $b\to s\,\,$ & $b\to d$ \\\hline
        Tree-dominated & 1 & $\bar \lambda^2$\\
        Annihilation-dominated & $\epsilon^3$ & $\bar\lambda^2\epsilon^3$\\\hline\hline
    \end{tabular}
    \qquad
    \begin{tabular}{l|c c}\hline\hline
        CP asymmetry & $b\to s\,\,$ & $b\to d$ \\\hline
        Tree-dominated & $\bar \lambda^2\epsilon^{2.5}$ & $\epsilon^{2.5}$\\
        Annihilation-dominated & $\bar \lambda^2\epsilon^{2}$ & $\epsilon^2$\\\hline\hline
    \end{tabular}
    \caption{Generic expectations for decay rates relative to the leading tree-dominated $b\to s$ decays (left) and for the CP asymmetries $A_{CP}$ and $\Delta S$ (right), according to our power counting.}
    \label{tab:power-counting-observables}
\end{table}

We start by checking the validity of the overall power counting for observables given in Table~\ref{tab:power-counting-observables}. Normalising to the tree-dominated $b\to s$ mode $\bar B^0\to D_s^-D^+$, we find that the measured rates correspond to our scaling usually within $30\%$, which is the expected range. We find slightly larger deviations of $\sim 40\%$ for $\bar B^0\to D^+D^-$ and $\bar B_s\to D_s^+D_s^-$, which is the first sign for negative interference of the sizeable annihilation amplitude $A_c$, as already emphasised in Ref.~\cite{Jung:2014jfa} and later confirmed in Ref.~\cite{Bel:2015wha}. This is in line with our assumed suppression for the annihilation topology. Most of the measured CP asymmetries $A_{CP}$ and $\Delta S_{CP}$ are consistent with zero so far, so it is difficult to infer their scaling. The new LHCb measurement for $A_{CP}(B^-\to D^0D^-)\sim2\%$ is slightly above 2 standard deviations away from zero and is on the lower side of our scaling $\epsilon^{2.5}\sim 5\%$, indicating that our treatment for the penguin amplitude is conservative.

\begin{figure}[t]
\includegraphics[width=0.5\textwidth]{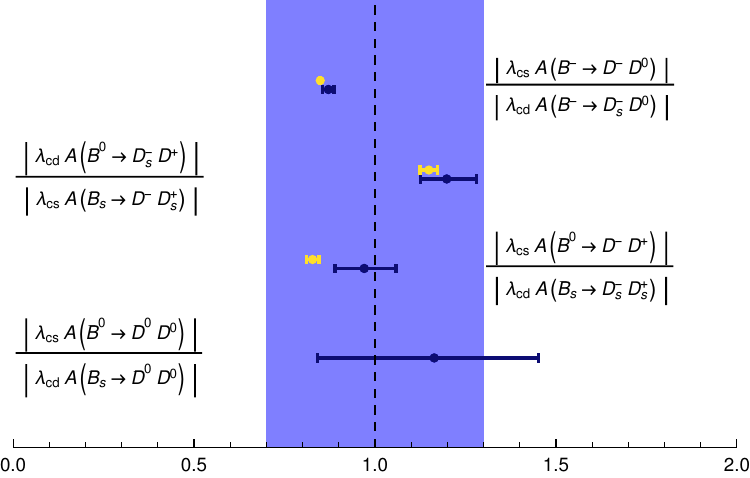}
\caption{Experimental determinations (dark blue data points) of ratios of absolute values of amplitudes compared to the theory expectation of $\mathcal{O}(\epsilon)$ $U$-spin breaking for a single amplitude~(blue band) and the theory prediction in the large $N_C$ limit (yellow data points) for decays with a contribution from a tree topology (see also the discussion in the text below.)\label{fig:plotUspinFact}
}
\end{figure}

In order to probe our assumption of approximate \SU{} symmetry, we consider the simple relations between so-called U-spin partners $\mathcal D_{b\to s}$ and $\mathcal D_{b\to d}$, that are related by interchanging all $d$ and $s$ quarks \cite{Fleischer:1999pa,Gronau:2000zy,Gronau:2000md},
\begin{align}
\frac{
    \Gamma_{\mathcal{D}_{b\rightarrow s}}
    }{
    \Gamma_{\mathcal{D}_{b\rightarrow d}}
    } &= 
    -\frac{
    A_{CP}^{\mathcal{D}_{b\rightarrow d}}
    }{
    A_{CP}^{\mathcal{D}_{b\rightarrow s}}
    }\,. \label{eq:U-spin-rule}
\end{align}
The overall smallness of $\mathcal A_u$, and thereby the direct CP asymmetries, renders these relations difficult to test. Instead, we can test U-spin symmetry using the relations

\begin{align}&
\left|\frac{\lambda_{cs}}{\lambda_{cd}}\right|^2\frac{
    \Gamma(B^-\rightarrow D^- D^0)
    }{
    \Gamma(B^-\rightarrow D_s^- D^0)
    } & \overset{\text{U-spin limit}}{\simeq}&
\left|\frac{\lambda_{cs}}{\lambda_{cd}}\right|^2\frac{
    \Gamma(\bar B_s\rightarrow D^- D_s^+)
    }{
    \Gamma(\bar B^0\rightarrow D_s^- D^+)
    }  
    &\overset{\text{U-spin limit}}{\simeq}&
\left|\frac{\lambda_{cs}}{\lambda_{cd}}\right|^2\frac{
    \Gamma(\bar B^0\rightarrow D^- D^+)
    }{
    \Gamma(\bar B_s\rightarrow D_s^- D_s^+)
    } & & \nonumber\\   && \overset{\text{U-spin limit}}{\simeq}& 1+\mathcal O(\epsilon^{2.5})\,, \label{eq:test-u-spin-1}\\
&\left|\frac{\lambda_{cs}}{\lambda_{cd}}\right|^2\frac{
    \Gamma(\bar B^0\rightarrow \bar{D}^0 D^0)
    }{
    \Gamma(\bar B_s\rightarrow \bar{D}^0 D^0)
    } &\overset{\text{U-spin limit}}{\simeq}
    &\left|\frac{\lambda_{cs}}{\lambda_{cd}}\right|^2\frac{
    \Gamma(\bar B^0\rightarrow D_s^+D_s^-)
    }{
    \Gamma(\bar{B}_s\rightarrow D^+ D^-)
    } &\overset{\text{U-spin limit}}{\simeq} & 1+\mathcal O(\epsilon^2)\,, \label{eq:test-u-spin-2}
\end{align}
following from Table~\ref{tab:topo_table}, since the \SU breaking in the leading amplitude $\mathcal A_c$ is, according to our power counting, much larger than the corrections due to the amplitudes $\mathcal A_u$ given above. 
We expect these \SU-breaking corrections generically at $\mathcal O(\epsilon)$, however at the \emph{amplitude level} for \emph{each mode}. In principle this allows for large corrections to these ratios, as emphasised in the context of $D$-meson decays \cite{Savage:1991wu,Hiller:2012xm}. 
The experimental values for the square root of these ratios are shown in Fig.~\ref{fig:plotUspinFact}. The fact that the ratios agree with unity within approximately $30\%$, indicated by the blue band, shows that the corrections at amplitude level actually tend to be smaller than the assigned size, again indicating that our treatment of this effect is conservative. These ratios confirm also the leading CKM structure of these decays.

In order to appraise the validity of the $1/N_C$ power counting, we overlay in Fig.~\ref{fig:plotUspinFact} also the prediction for the respective ratios in the $N_C\rightarrow \infty$ limit. We neglect all contributions except for the tree diagram, since they are either strongly suppressed or not fully factorisable even in the large-$N_C$ limit. In this limit, the tree diagrams factorise and can be calculated using the $D_{(s)}$ meson decay constant and the $B_{(s)}\rightarrow D_{(s)}$ form factor. We stress again that these decays do not factorise in the heavy-quark limit~\cite{Beneke:1999br,Beneke:2000ry}. The factorised results show the well-known unphysical scale-dependence at leading order (which is, however, relatively small for the tree amplitude), and receive corrections at the level of $1/N_C^2$, not included in the error bars shown.
In order to evaluate the factorisation formula, we use the results 
shown in Table~\ref{tab:form-factors}.
\begin{table}[]
    \centering
    \begin{tabular}{ccc}
    \hline \hline
      Decay  & $f_{D_{(s)}}$ & $f_0(\overline{m_D}^2)$\\
      \hline
       $B \to D$  & $(212.0 \pm 0.7)$ MeV &
       $0.726\pm0.011$\\
       $B_s \to D_s$ & $(249.9 \pm 0.5)$ MeV &
       $0.729\pm0.011$ \\
       \hline \hline
    \end{tabular}
    \caption{Decay constants and form factors for both $B \to D$ and $B_s \to D_s$ decays, using data from \cite{FlavourLatticeAveragingGroupFLAG:2021npn, Bazavov:2017lyh, Carrasco:2014poa} for the decay constants and from a BGL fit to Refs.~\cite{Na:2015kha,MILC:2015uhg,McLean:2019qcx} for the form factors. We use $\overline{m_D}^2= 1/3 \times (m_{D^0}^2 + m_{D^-}^2 + m_{D^-_s}^2)$.
    }
    \label{tab:form-factors}
\end{table}

From Fig.~\ref{fig:plotUspinFact} we see that, overall, the $1/N_C$ limit works well for the $U$-spin ratios.
We interpret the result for the ratio $\vert \lambda_{cs} A(B^0\rightarrow D^-D^+) \vert / \vert \lambda_{cd} A(B_s\rightarrow D_s^- D_s^+)\vert$ again as showing the importance of annihilation contributions, which also display a different \SU-breaking pattern from the tree amplitude. We also checked that the individual tree-dominated modes are reasonably well described in the large-$N_C$ limit, not just the $U$-spin ratios. In this case, the observed rates are consistently smaller than those obtained from our fit in Table ~\ref{tab:BRsJD}, with corrections of $17-28\%$ on amplitude level for the central values of the modes without annihilation contributions, which is on the large side for $1/N_C^2$ corrections. 
This pattern is reminiscent of the one found in Ref.~\cite{Bordone:2020gao} in $b\to c\bar u(d,s)$ transitions, however in the case of \BDD{} the discrepancy with theory expectations is not as significant, since QCD factorisation is not applicable, as emphasised before.  
The overall agreement of the $1/N_C$ limit with the data serves as motivation to employ the $1/N_C$ counting as introduced in Sec.~\ref{sec:power-counting}, which is in line with the above observations. We stress that we do not employ the $N_C\rightarrow \infty$ limit in our global fit, it serves here only as a cross-check.

\subsection{Key observables}

Having tested the ingredients to our power counting, we can proceed by analysing its phenomenological consequences. Below we perform a global fit, which employs all available information simultaneously and hence allows for maximal control over the subleading amplitudes discussed above. However, it is still useful to consider the most important observables separately, in order to understand what kinds of contributions we are trying to control and what the main physics results of the global analysis will be, apart from a better understanding of \BDD{} decays.

\subsubsection{Time-dependent CP asymmetry in $\bar B_s\to D_s^+D_s^-$}
This observable is the main reason for the interest in \BDD{} decays: it offers a determination of the $B_s$ mixing phase that is as clean as that in $\bar B_s\to J/\psi\phi$. In the limit of vanishing $\mathcal A_u$, the relations
\begin{align}
    S_{CP}(\bar B_s\to D_s^-D_s^+) = -\sin(\phi_s)\qquad \mbox{and}\qquad A_{CP}(\bar B_s\to D_s^-D_s^+) = 0 \label{eq:SCPBsDsDs}
\end{align}
hold and corrections to these relations can be read off from Table~\ref{tab:topo_table} (see also Table~\ref{tab:power-counting-observables}), the largest contribution being
\begin{align}
    \Delta S(\bar B_s\to D_s^-D_s^+)\sim \bar\lambda^2\epsilon^{2.5}\lesssim 1\%\,. \label{eq:DeltaSCPBsDsDs}
\end{align}
Importantly, even this small pollution can be controlled within the global fit, in particular with the help of the related $b\to d$ mode $\bar B^0\to D^-D^+$ \cite{Fleischer:1999nz,Fleischer:2007zn,Gronau:2008ed,Jung:2014jfa,Bel:2015wha}. Using the mixing phase $\phi_d$ as an external input, and observing the smallness of $\phi_s$, the approximate formula
\begin{align}
    \sin(\phi_s)  = -\left[S_{CP}(\bar B_s\to D_s^-D_s^+) +\frac{\bar\lambda^2}{\cos(\phi_d)}\left(S_{CP}(\bar B^0\to D^-D^+)+\sin(\phi_d)\right)\right]+\mathcal O(\bar \lambda^2\epsilon^{3.5}) \label{eq:sinphis}
\end{align}
holds, which is similar to a relation between $\bar B^0\to J/\psi K_S$ and $\bar B^0\to J/\psi \pi^0$ \cite{Jung:2012mp} (see also \cite{Ligeti:2015yma}) and is only complicated by the different mixing phases for the two modes involved. 
The relation Eq.~(\ref{eq:sinphis}), as well as part of the \SU-breaking corrections to it, are included automatically in the global fit. 

\subsubsection{CP asymmetries in other $b\to s$ transitions}
The statements made for $\bar B_s\to D_s^-D_s^+$ hold similarly for the CP asymmetries in other $b\to s$ transitions (see Table~\ref{tab:power-counting-observables}). Specifically, there is no test yet of any CP asymmetry in the annihilation-dominated modes, which constitute new-physics tests complementary to the tree-dominated modes.

\subsubsection{Quasi-isospin relations}
We define quasi-isospin relations as those which require, in addition to pure isospin symmetry, some weak assumption about the scaling of suppressed amplitudes, which holds, however, to a level comparable to that of isospin itself. Almost all relations between just two modes fall within this class, for instance also between $B\to J/\psi K^0$ and $B\to J/\psi K^+$. In the case of \BDD{}, the relations for observables discussed in Ref.~\cite{Jung:2014jfa} (see Ref.~\cite{Gronau:1995hm} for earlier partial results) read, with the power counting adopted in this work, as
\begin{align}\label{eq:quasiiso}
    \Gamma(\bar B^0\to D_s^- D^+) = \Gamma(B^-\to D_s^-D^0)(1+\mathcal O(\bar\lambda^2 \epsilon^{2.5}))\quad \mbox{and}\quad\Gamma(\bar B_s\to \bar D^0 D^0) = \Gamma(\bar B_s\to D^-D^+)(1+\mathcal O(\bar\lambda^2 \epsilon^2))\,,
\end{align}
up to isospin-breaking contributions not included here. When these rates are determined using precise information on the production fractions and detection efficiencies from elsewhere, they can be used to test for BSM physics with $\Delta I=1$, or, assuming the absence of such contributions, to investigate isospin breaking in the SM. On the other hand, they can be used to determine the relative production fractions of neutral and charged $B$ mesons, \emph{i.e.}, $f_d/f_u$ at the LHC and the relative decay rates of $\Upsilon(4S)$ to $\bar B^0B^0$ and $B^+B^-$ at Belle~(II), in case of the first relation. The second relation can also be used to validate the detection efficiencies of charged vs.~neutral $D$ mesons, since there the initial state is identical. Again, these relations are included automatically in the global fit, and benefit from additional information, for instance the presence of the production fractions and relative decay rates in other observables in the fit.

\subsection{Global Fit and Predictions \label{sec:global-fit}}

We perform a global fit to all available experimental data, given in Tables~\ref{tab:CPasym-exp},~\ref{tab:CKMinput},~\ref{tab:br-input-data}, and~\ref{tab:new_inputs}, using the setup developed in Section~\ref{sec:power-counting}. This enables us to go beyond tests of just the relations for individual decay modes discussed above: 
their combination determines the symmetry and symmetry-breaking structure of \BDD{} decays in general. This structure,
together with our power counting, induces non-trivial correlations on the theory side, allowing the prediction of so far unmeasured observables. The global fit also increases the overall sensitivity and allows for checking the consistency of not only the listed relations, but also their breaking.

For the moment, the measurement of the time-dependent CP asymmetry in $\bar B_s\to D_s^-D_s^+$ does not constrain the weak mixing phase $\phi_s$ at a level competitive with its indirect determination in global fits \cite{CKM21,UTfit:2022hsi} or its measurement in $B_s\to J/\psi \phi$ \cite{LHCb:2023sim}. We therefore use this value as external input in our fit, in order to be able to determine the variations around this value caused by subleading contributions. Once more precise measurements in \BDD{} are available, we will remove the external input and use our fit to determine $\phi_s$ directly instead. On the technical side, we employ a $\chi^2$ fit, extracting the topological amplitudes introduced in Sec.~\ref{sec:power-counting} from the experimental data. The bounds on the parameters are implemented using the Rfit approach \cite{Hocker:2001xe}. 

\begin{table}[t]
    \centering
\begin{tabular}{ccccc}
    \hline \hline
        & Observable & Global fit  & Exp.-only fit  & PDG \cite{PDG2022} \\
        \hline 
         1 & $\mathcal{B}(B^- \to D^- D^0)$ & 
         $0.366\pm0.025$ & 
         $0.383^{+0.034}_{-0.033}$ & $0.38\pm0.04$\\
         2 & $\mathcal{B}(B^- \to D_s^- D^0)$  & 
         $8.8\pm0.6$ & 
         $9.2^{+0.9}_{-0.8}$ & $9.0\pm0.9$\\
         3 & $\mathcal{B}(\bar{B}^0 \to D_s^- D^+)$   & 
         $8.2\pm0.5$ & 
         $7.6\pm0.7$ & $7.2\pm 0.8$\\
         4 & $\mathcal{B}(\bar{B}_s \to D^- D_s^+)$  & 
         $0.303^{+0.045}_{-0.040}$ &
         $0.280^{+0.047}_{-0.042}$ & $0.28\pm0.05$\\
         5 & $\mathcal{B}(\bar{B}^0 \to D^- D^+)$   & 
         $0.224^{+0.022}_{-0.021}$ & 
         $0.231^{+0.023}_{-0.022}$ & $0.211\pm 0.018$\\
         6 & $\mathcal{B}(\bar{B}_s \to D_s^- D_s^+)$   & 
         $4.7\pm0.6$ & 
         $4.4\pm0.6$ & $4.4\pm0.5$\\
         7 & $\mathcal{B}(\bar{B}^0 \to D_s^- D_s^+)$   & 
         $0.024^{+0.013}_{-0.020}$ & $\leq 0.036$ & $\leq 0.036$\\
         8 & $\mathcal{B}(\bar{B}_s \to D^- D^+)$   & 
         $0.204^{+0.036}_{-0.033}$ &
         $0.24_{-0.05}^{+0.06}$ & $0.22 \pm 0.06$\\
         9 & $\mathcal{B}(\bar{B}^0 \to \bar{D}^0 D^0)$   & 
         $0.013\pm0.006$&
         $0.012\pm0.006$ & $0.014\pm0.007$\\
         10 & $\mathcal{B}(\bar{B}_s \to \bar{D}^0 D^0)$   & 
         $0.203^{+0.036}_{-0.033}$ &
         $0.166^{+0.042}_{-0.037}$  
         & $0.19\pm 0.05$\\
         \hline \hline
    \end{tabular}
    \caption{Branching ratios in units of $10^{-3}$ resulting from our global fit (second column), compared to our experiment-only fit (see Sec.~\ref{sec:exp}, third column) and the measurements as quoted in the PDG~\cite{PDG2022} (fourth column). \label{tab:BRs-theory}}
\end{table}

\begin{figure}[t]
    \centering
    \subfigure[$b \to s$ $T$-dominated]{\includegraphics[height=4.6cm]{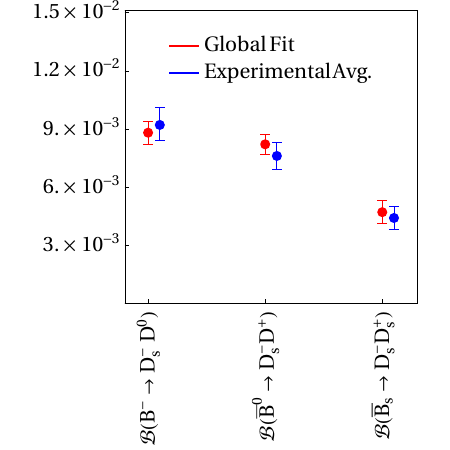}}\hfill
    \subfigure[$b \to d$ $T$- and $b \to s$ $A^c$-dominated]{\includegraphics[height=4.6cm]{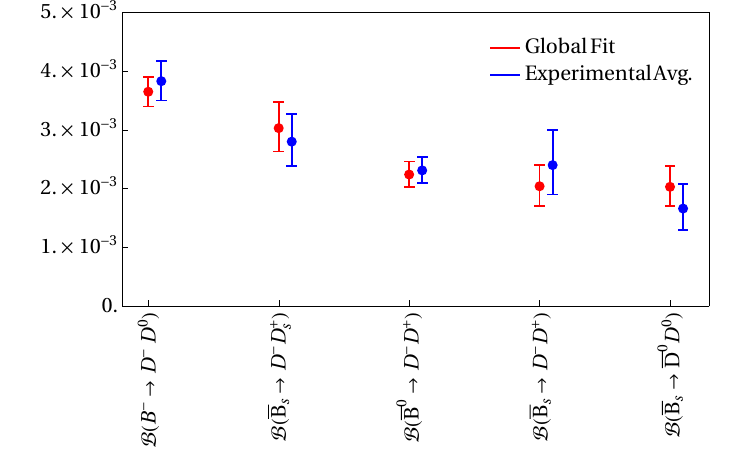}}\hfill
    \subfigure[$b \to d$ $A^c$-dominated]{\includegraphics[height=4.6cm]{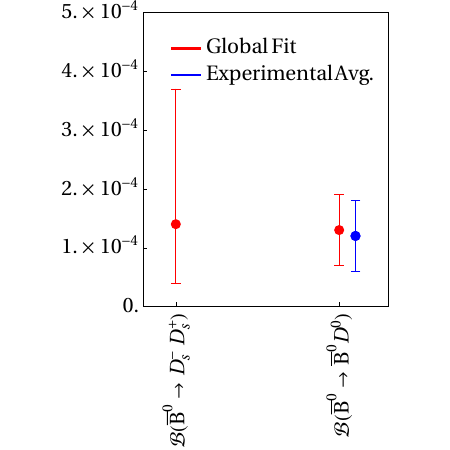}}
    \caption{Comparison of branching ratios extracted from global fit to those obtained from our experiment-only fit, as seen in Table~\ref{tab:BRs-theory}
    \label{fig:plotBR}}
\end{figure}

With this fit setup, we obtain an excellent description of the input observables, with all of them reproduced within the $2\sigma$ range. None of the parameters has a relevant pull towards outside the range imposed by our power counting and this is another confirmation of our conservatively-chosen power counting (see also the discussion above). The na\"{i}vely evaluated fit quality is good, with $\chi^2_{\mathrm{min}}=8.6$ for $43-35=8$ degrees of freedom, corresponding to a confidence level of $37\%$. This value is, however, not very meaningful in our approach: as emphasised before, we include a number of amplitudes only in order to be able to give conservative uncertainty estimates. For instance, the experimental uncertainties for the annihilation-dominated modes are too sizeable to pin down the \SU{}-breaking in the leading $A_c$ amplitude. Setting this breaking to zero, however, would yield unrealistically small uncertainties for the branching fractions predicted by the fit. Overall, we include 10 parameters in our fit that are barely constrained beyond our power counting. Removing all of them increases the minimal $\chi^2$ only by 0.7
and evaluating the confidence level in this setup would yield $95\%$. In fact, presently only very few parameters are, strictly speaking, necessary to describe all available measurements: the leading tree amplitude obviously, the sizeable annihilation amplitude discussed previously, 
the \SU-breaking contributions to the leading amplitude, 
and a contribution to $A_{CP}(B^-\to D^-D^0)$,
none of them even close to the limit dictated by our power counting.
While with more precise measurements, more hadronic parameters will become necessary in the fit, presently all but five of them are only included in order to obtain conservative predictions for the so far unmeasured observables.

The fit results for the branching fractions are given in Table~\ref{tab:BRs-theory} and illustrated in Fig.~\ref{fig:plotBR}. We make the following observations:
\begin{enumerate}
    \item The fit results for all branching fractions are compatible with the values obtained in the fit without imposing the \SU{} structure. Due to that structure, however, the branching fractions are generally more precisely determined in the global fit. 
    \item The two quasi-isospin relations encoded in the fit have different effects. The relation on the left in Eq.~\eqref{eq:quasiiso} mostly leads to the determination of the ratio of production fractions: the measured value of the ratio
    \begin{align}
        \frac{\Gamma_{\overline{B}^0\rightarrow D_s^- D^+}}{\Gamma_{B^-\rightarrow D_s^- D^0}} &= 0.88\pm0.05
        \label{eq:isospin-1}
    \end{align}
    when setting $f_d/f_u$ to unity (c.f.~Table~\ref{tab:BRsJD}) translates in the global fit to
    \begin{align}
        (f_d/f_u)^{\text{LHCb,~7 TeV}} &=   \label{eq:fd-over-fu-th} 0.86\pm0.05\,.
    \end{align} 
    This value is surprisingly far away from its na\"{i}ve expectation of unity, with a significance of $2.5\sigma$. Note, that this tension cannot be remedied by simply allowing for a larger amplitude $A_1^u$: removing the constraints from our power counting for this parameter can only reduce the tension to $\sim 1.6\sigma$, even with $A_1^u$ of the same order as the leading tree amplitude. If this value for $f_d/f_u$ is confirmed, it would have significant consequences for several branching fractions determined at LHCb involving this fraction, for instance $\mathcal{B}(\bar B_{(s)}\to\mu^+\mu^-)$, which uses $B^+\to J/\psi K^+$ as normalisation mode and applies $f_s/f_d$ assuming $f_d=f_u$ \cite{CMS:2014xfa,LHCb:2021awg}. It should therefore be measured using additional data as well as other decay modes, like $\Gamma( B^0\to J/\psi  K^0)/\Gamma(B^+\to J/\psi K^+)$, and compared to precision determinations at Belle(-II) to differentiate the cases $f_d/f_u\neq 1$ and $\Gamma(\bar B^0\to D_s^-D^+)/\Gamma(B^{-}\to D_s^-D^0)\neq 1$. The same holds for the determination of $\mathcal{B}(\bar{B}_{(s)}\rightarrow \mu^+\mu^-)$ by the CMS collaboration \cite{CMS:2014xfa,CMS:2022mgd}, however the value of $f_d/f_u$ must be assumed to be different from that at LHCb, given the $p_T$-dependence of the other production fractions. The CMS measurement of $f_d/f_u$ in Ref.~\cite{CMS:2022wkk} is compatible with unity.
    The relation on the right of Eq.~\eqref{eq:quasiiso}, on the other hand, effectively implies an average of the two branching fractions involved. The central value of the corresponding ratio, 
    \begin{align}
        \frac{\Gamma_{\overline{B}_s\rightarrow \overline{D}^0 D^0}}{\Gamma_{\overline{B}_s\rightarrow D^-D^+}} =  0.69^{+0.25}_{-0.18}
        \label{eq:isospin-2} 
    \end{align}
    is far away from unity, but only with a significance of about $1\sigma$. 
    \item 
    The quasi-isospin relations above stem from the fact that the $\Delta I=1$ contributions to \BDD{} are heavily suppressed in $b\to s$ transitions. The same is true for the $\Delta I=3/2$ contributions in $b\to d$ transitions, however in this case the additional CKM suppression by $\bar\lambda^2$ is absent \cite{Jung:2014jfa}. The resulting rate sum rule has therefore larger corrections:
    \begin{align}
        \frac{\Gamma(\bar{B}^0\rightarrow D^-D^+)}{\Gamma(B^-\rightarrow D^-D^0)} &= 1 + \mathcal{O}(\epsilon^{3/2}) = 0.65^{+0.09}_{-0.08}\,. \label{eq:sum-rules-1}
    \end{align}
    This sizeable shift is approximately given by $2 \mathrm{Re}(A_c)/T$ and reflects again the sizeable annihilation amplitude $\vert \mathrm{Re}(A_c)/T\vert \sim 18\%$ noted before (see also Ref.~\cite{Jung:2014jfa}).
    \item Given the range for the $U$-spin ratios in Fig.~\ref{fig:plotUspinFact} and the absence of tensions in the fit, the fitted $U$-spin ratios are similar to their measured values obtained without imposing the \SU{} structure. 
    \item 
    The one branching fraction in \BDD{} that has not yet been significantly measured, $\mathcal B(\bar B^0\to D_s^-D_s^+)$, is predicted by our fit to lie in the interval
    \begin{align}
        \mathcal B(\bar B^0\to D_s^-D_s^+) \in [0.004,0.037]\, \times 10^{-3}\quad(68\% \mathrm{CL})\,,
    \end{align}
    in accordance with the existing upper limit $\mathcal B(\bar B^0\to D_s^-D_s^+)\leq 0.036\times10^{-3}$, which was not used in the fit.
\end{enumerate}
These observations show the importance of improving on the measurements of the \BDD{} branching fractions. Prospects for this are excellent, since very few of the existing measurements used the full currently available data set of the corresponding experiment, and the Belle~II dataset has not yet been exploited for \BDD{} measurements.\\

\begin{table}[t]
    \centering
    \begin{tabular}{l|ccc|ccc}
    \hline \hline
        Mode & \multicolumn{3}{c|}{$A_{CP} (\%) $}  & \multicolumn{3}{c}{$S_{CP} (\%)$} \\\hline
             &  global fit &  w/o CP inputs  & exp. &  global fit & w/o CP inputs    & exp.      \\
        \hline 
         $B^- \to D^- D^0$ &
         [1.2,3.1] & $[-34.0,34.0]$ & $2.4\pm 1.1$ &-- &-- &-- \\
         $B^- \to D_s^- D^0$ & 
         [-0.2, 0.3] & $[-1.6,1.6]$ & $0.5\pm0.6$ &  --& -- &--\\
         $\bar{B}^0 \to D_s^- D^+$ & 
         [-0.8, 0.6] & $[-0.8, 0.8]$ & &  -- &-- &--\\
         $\bar{B}_s \to D^- D_s^+$ & 
         [-9.3, 18.1] & $[-18.6, 18.6]$ & &  --& -- &--\\
         $\bar{B}^0 \to D^- D^+$ &
         [-3.5, 25.3] & $[-28.2, 28.2]$ & $13\pm17$ &  [-87.2, -60.4]& $[-87.9, -46.7]$ & $-81\pm20$\\
         $\bar{B}_s \to D_s^- D_s^+$ & 
         [-1.3, 0.7] & $[-1.4, 1.4]$ & $-8\pm 18$ & [2.9, 5.0] & $[2.3,5.0]$ & $-2\pm 17$\\
         $\bar{B}^0 \to D_s^- D_s^+$ &
         [-43.0,45.1] & [-60.4, 60.4] & &  [-94.2, -31.0] & $[-99.0, -15.1]$ & \\
         $\bar{B}_s \to D^- D^+$ &
         [-1.5, 1.5] & $[-1.6, 1.6]$& & [2.1, 5.2] & $[2.1,5.3]$ & \\
         $\bar{B}^0 \to \bar{D}^0 D^0$ &
         [-62.7,61.3] & $[-64.7, 64.7]$ & &  [-99.3,-11.7] & [-99.5,-8.1] & \\
         $\bar{B}_s \to \bar{D}^0 D^0$ &
         [-3.1,3.1] & $[-3.3, 3.3]$ & &  [0.5,6.8] & $[0.4,6.9]$ & \\
         \hline \hline
    \end{tabular}
    \caption{Predictions for $A_{CP}(\mathcal{D})$ and $S_{CP}(\mathcal{D})$ resulting from our global fit (columns two and five) and when removing all CP asymmetry input data from the fit (columns three and six), compared to the available experimental measurements. We quote the 68\% CL regions and give no central points for the fitted values as we observe sizeable flat directions in the fit for several of the CP asymmetries.}
    \label{tab:CP-asymmetries}
\end{table}

\begin{figure}[t]
  \centering
  \subfigure[\, $b\rightarrow d$ transitions.]{\includegraphics[width=8cm]{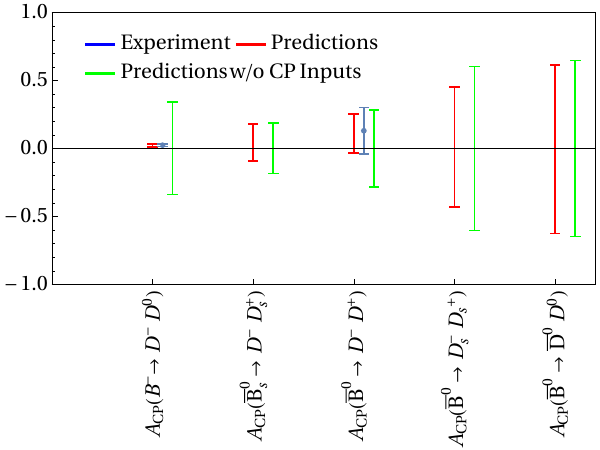}}\hspace{1em}%
  \subfigure[\, $b\rightarrow s$ transitions.]{\includegraphics[width=8cm]{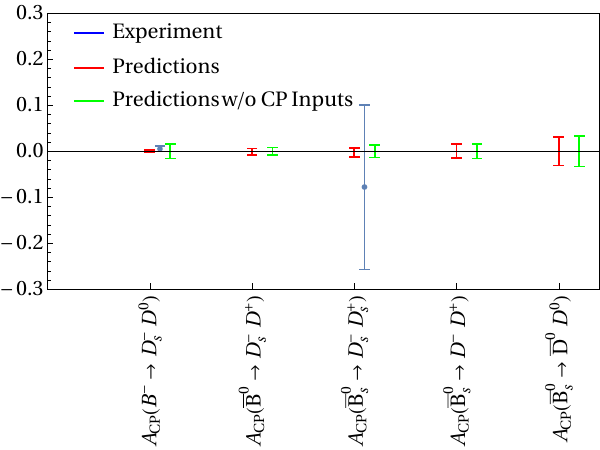}}
 \caption{Predictions for $A_{CP}(\mathcal{D})$ for the different channels resulting from our global fit (red) compared with the fit when removing all CP asymmetry measurements (green) and the experimental results (blue) when available.
 We show the 68\% CL regions and give no central points as we observe sizeable flat directions in the fit for several of the CP asymmetries. The  value shown for $A_{CP}(B^-\rightarrow D^-D^0)$ is the weighted average of the two corresponding ones given in Table~\ref{tab:CPasym-exp}. \label{fig:1D-ACPs}}
\end{figure}

\begin{figure}[t]
  \centering
  \subfigure[\, $b\rightarrow d$ transitions.]{\includegraphics[width=8cm]{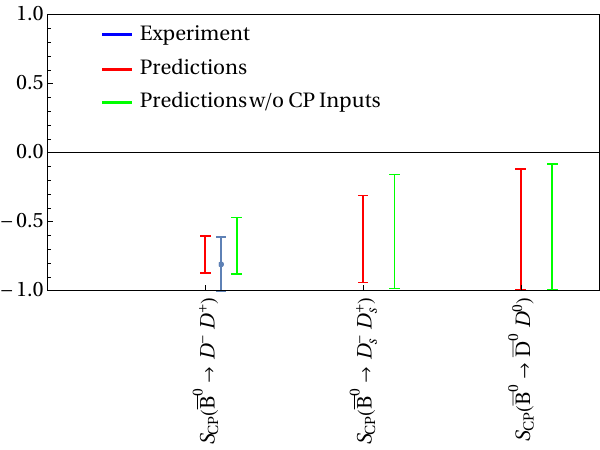}}\hspace{1em}%
  \subfigure[\, $b\rightarrow s$ transitions.]{\includegraphics[width=8cm]{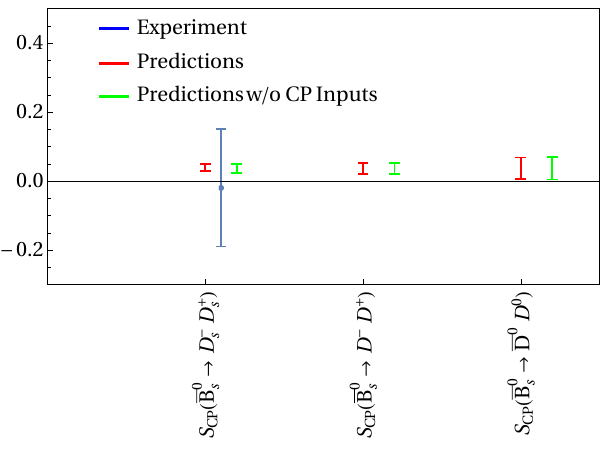}}
 \caption{Predictions for $S_{CP}(\mathcal{D})$ for the different channels resulting from our global fit (red) compared with the fit when removing all CP asymmetry measurements (green) and the experimental results (blue) when available. 
  We show the 68\% CL regions and give no central points as we observe sizeable flat directions in the fit for several of the CP asymmetries.
  \label{fig:1D-SCPs}}
\end{figure}

\begin{figure}
\begin{center}
\subfigure[\label{fig:ACP1_ACP3_2D}]{\includegraphics[width=0.32\textwidth]{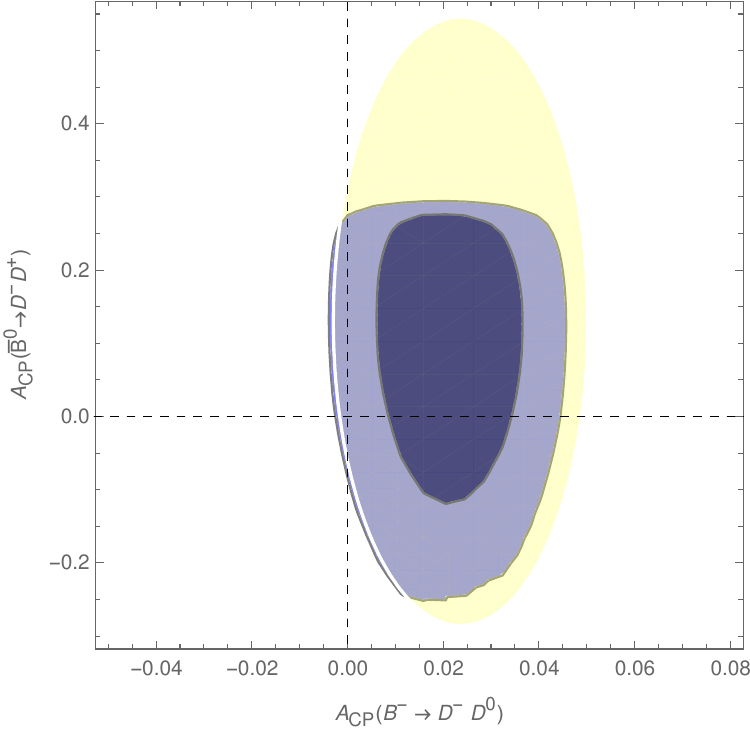}}
\subfigure[\label{fig:ACP1_ACP6_2D}]{\includegraphics[width=0.32\textwidth]{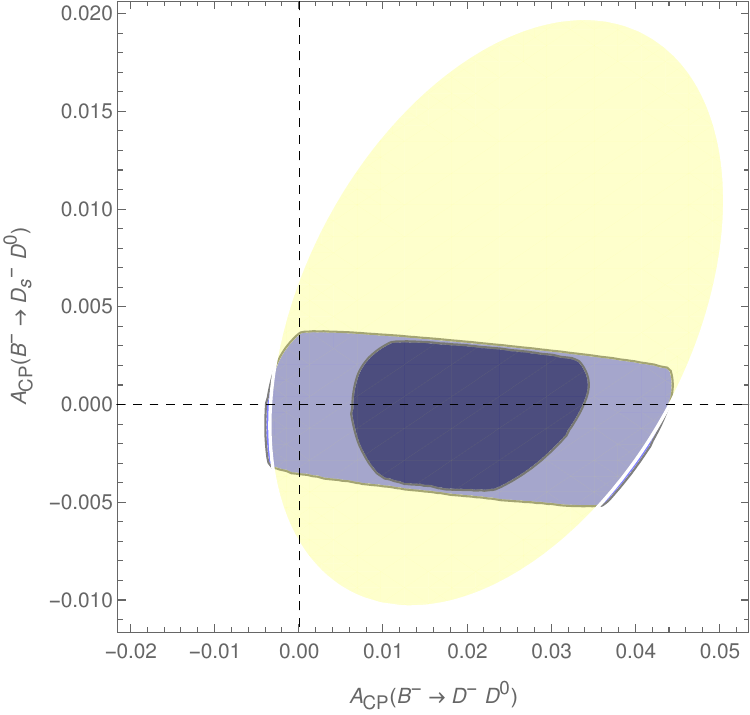}}
\subfigure[\label{fig:SCP3_SCP8_2D}]{\includegraphics[width=0.32\textwidth]{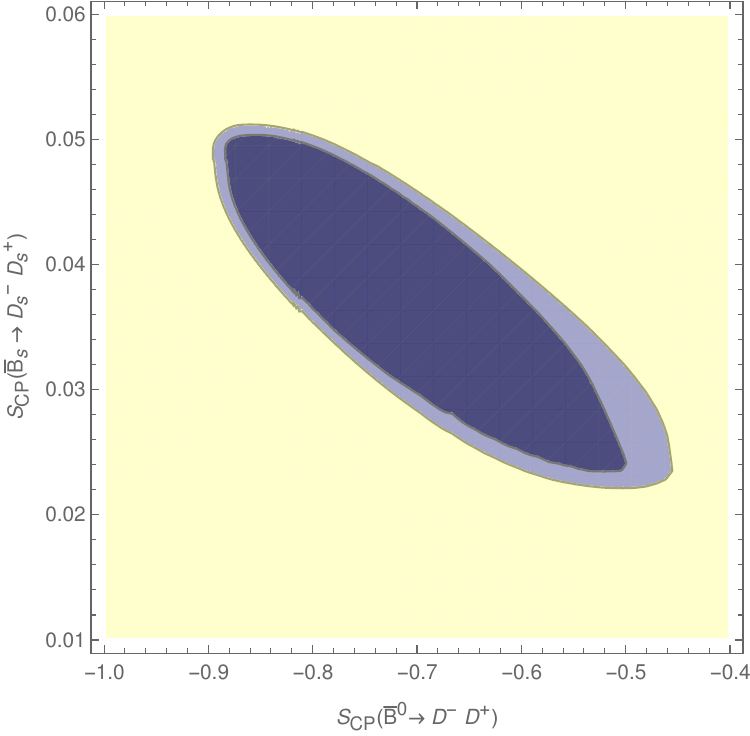}}
\subfigure[\label{fig:ACP3_SCP3_2D}]{\includegraphics[width=0.32\textwidth]{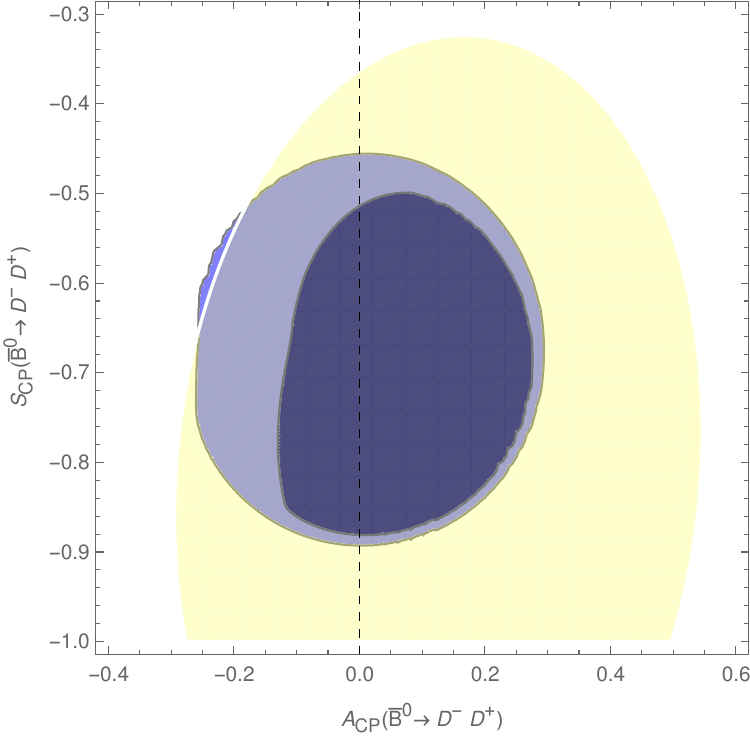}}
\subfigure[\label{fig:ACP4_SCP4_2D}]{    \includegraphics[width=0.32\textwidth]{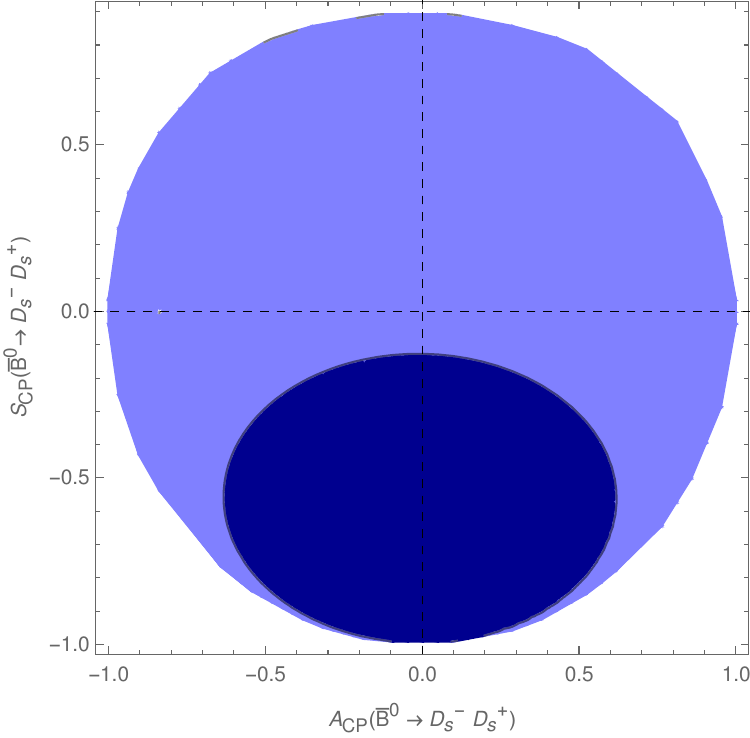}}
\subfigure[\label{fig:ACP5_SCP5_2D}]{    \includegraphics[width=0.32\textwidth]{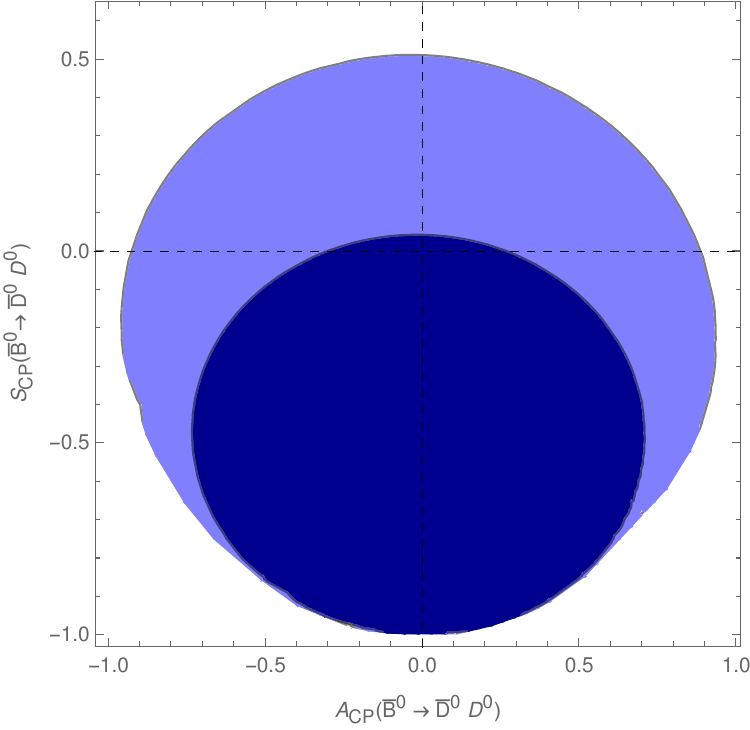}}
\subfigure[\label{fig:ACP8_SCP8_2D_Meas}]{    \includegraphics[width=0.32\textwidth]{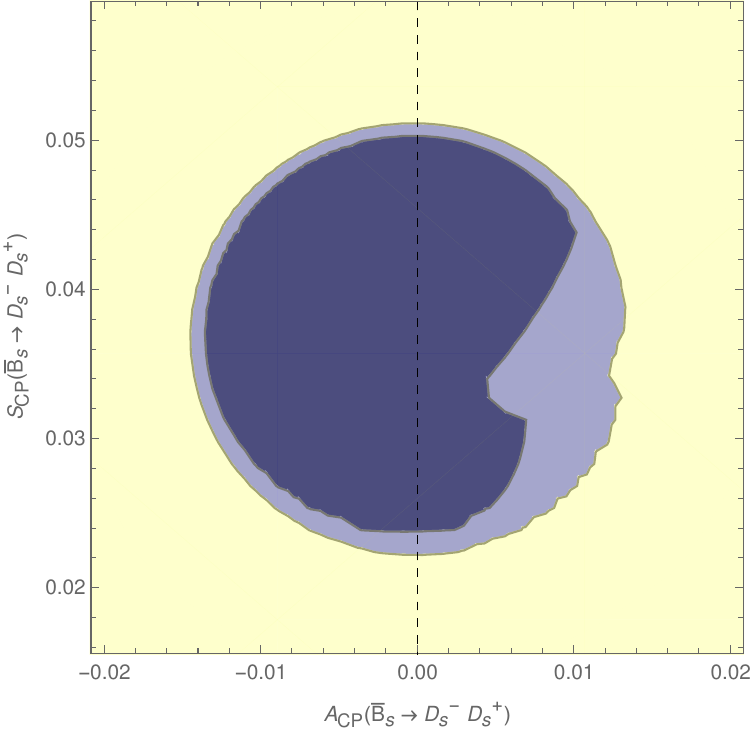}}
 \subfigure[\label{fig:ACP9_SCP9_2D}]{    \includegraphics[width=0.32\textwidth]{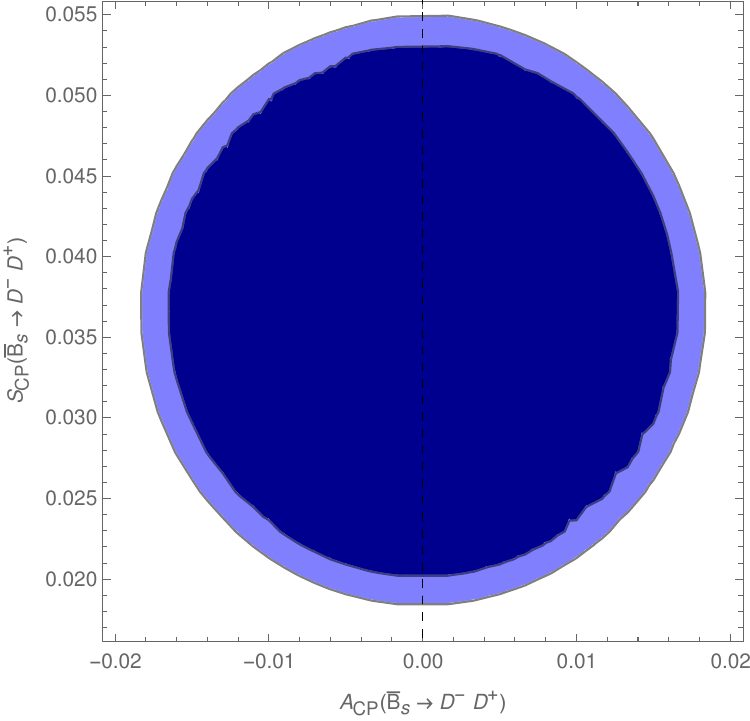}}
\subfigure[\label{fig:ACP10_SCP10_2D}]{    \includegraphics[width=0.32\textwidth]{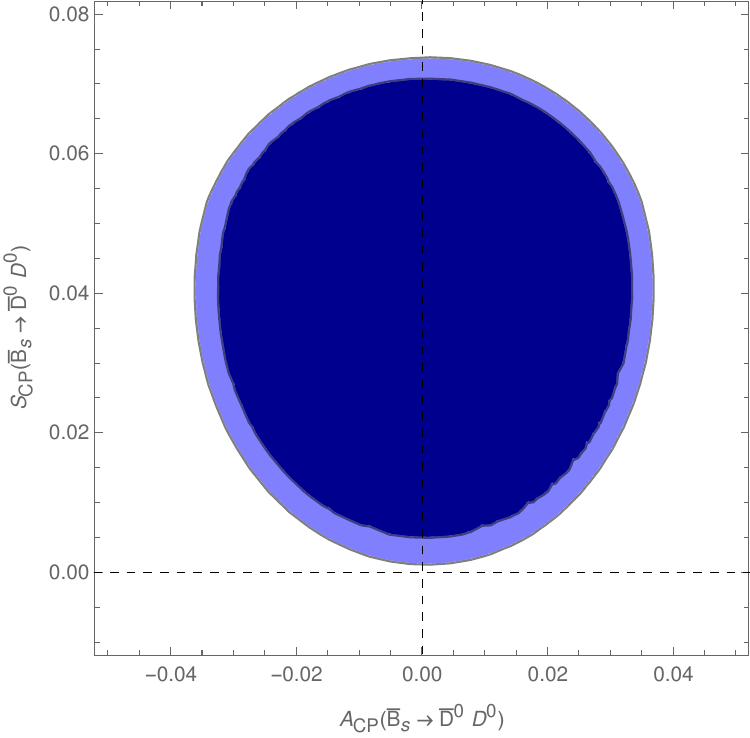}}
    \end{center}
    \caption{Theory predictions for CP asymmetries from the global fit at 68\% CL ($\Delta\chi^2 = 2.30$, dark blue) and 95\% CL ($\Delta\chi^2=5.99$ light blue), compared to the experimental data (where available) at 95\% CL (light yellow). The  experimental result shown for $A_{CP}(B^-\rightarrow D^-D^0)$ is the weighted average of the two corresponding ones given in Table~\ref{tab:CPasym-exp}.}\label{fig:2D-plots}
\end{figure}

\begin{figure}
    \centering
    \includegraphics[width=0.5\textwidth]{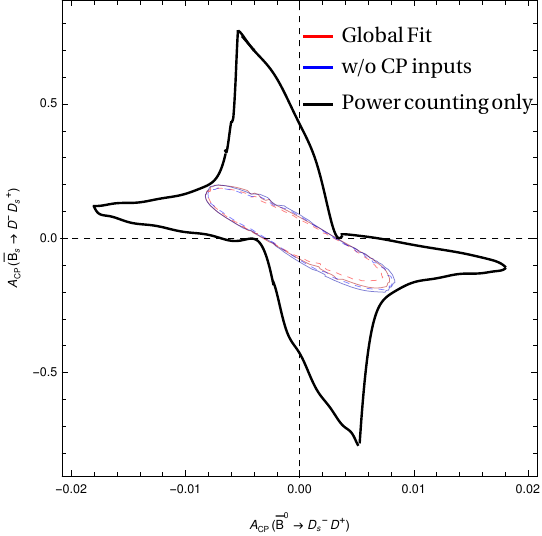}
    \caption{Correlation between $A_{CP}(\overline{B}^0\rightarrow D_s^-D^+) $ and $A_{CP}(\overline{B}_s\rightarrow D^- D_s^+)$
    as illustration of the impact of different experimental inputs on the fit. We show the global fit result at 68\%~CL~(red-dashed) and 95\%~CL~(red-solid), the fit without CP inputs at 68\% CL~(blue-dashed) and 95\%~CL~(blue-solid) as well as the constraints from power counting only, without any experimental inputs~(black-solid).}
    \label{fig:ACP2_ACP7_Combined}
\end{figure}

The fit results for the CP asymmetries are given in Table~\ref{tab:CP-asymmetries} and illustrated in Figs.~\ref{fig:1D-ACPs}, \ref{fig:1D-SCPs}, \ref{fig:2D-plots}, and \ref{fig:ACP2_ACP7_Combined}. The strength of our methodology lies in the prediction of CP asymmetries through their correlation to other observables by the \SU{} analysis. As many of the CP asymmetries have not yet been measured, the information on several topologies whose contributions to the branching ratios are suppressed mainly comes, for now, from the power counting introduced in Sec.~\ref{sec:power-counting}. This causes flat directions in the fit for several of the CP asymmetries; we give $\Delta \chi^2=1$-intervals in Table~\ref{tab:CP-asymmetries} in order to avoid the impression that these values were corresponding to Gaussian distributions. As more measurements become available, and with increased precision, the results of our global fit for the as yet unmeasured CP asymmetries will also become more precise.
We make the following observations:
\begin{enumerate}
    \item The available experimental data are well described in the fit; there are no tensions. 
    \item Figs.~\ref{fig:1D-ACPs}, \ref{fig:1D-SCPs} visualise how specifically the CP asymmetries measured in charged $B$ decays and $\bar B^0\to D^-D^+$ constrain the fit significantly. Thanks to these measurements, the global fit already constrains most ranges for CP asymmetries beyond our power counting, as illustrated by comparing the red and green ranges.
    $A_{CP}(B^-\to D^-D^0)$ in particular illustrates three things. Firstly, our power counting is extremely conservative, implying that measuring asymmetries outside our given ranges would indicate BSM physics, not underestimated hadronic contributions. Secondly, it illustrates the strength of the mechanism we use to constrain penguin pollution. This is seen from $A_{CP}(B^-\to D_s^-D^0)$, which, despite having been measured more precisely than our prediction from power counting alone, is significantly more strongly constrained in the global fit  than by its measurement, due to its correlation with $B^-\to D^-D^0$, shown in Fig.~\ref{fig:2D-plots}(b). This is a generic feature of $b\to d$ vs. $b\to s$ transitions, applying in particular to all $U$-spin partners, but also other \SU{} partners. Thirdly, it shows the necessity to measure the CP asymmetries in more modes. This becomes clear when considering the sizeable range for $A_{CP}(\bar B^0\to D^-D^+)$ still allowed in the global fit. While it shares its main contribution with $B^-\to D^-D^0$, which is measured to $\sim 1\%$, the presence of the second contribution of the same size in the latter means that the correlation is weakened, due to possible cancellations, as seen comparing plots (a) and (b) in Fig.~\ref{fig:2D-plots}. A measurement of the CP asymmetry in $\bar B_s\to D^-D_s^+$, for instance, would allow these contributions to be constrained separately, improving the predictions for other modes.
    \item The precision of the prediction for the CP asymmetries in $\bar B_s\to D_s^-D_s^+$ means that this mode can be used to measure $\phi_s$ with negligible theory uncertainties for the foreseeable future, and even the small penguin pollution can be controlled in our fit.
    \item Fig.~\ref{fig:ACP2_ACP7_Combined} illustrates the non-trivial interplay between the experimental data and the input from theory within the global fit. We compare the \lq\lq{}plain theory\rq\rq{} prediction with no experimental inputs at all, following from the underlying parametrisation and the power counting alone, with that using different experimental inputs. We find that already the measured branching fractions constrain the CP asymmetries significantly, by fixing their denominators, and the available measurements of CP asymmetries constrain them further, as noted already above.
    \item The two-dimensional correlation plots in Fig.~\ref{fig:2D-plots} and ranges in Figs.~\ref{fig:1D-ACPs}, \ref{fig:1D-SCPs} demonstrate again more generally how much parameter space is left to explore in order to test for BSM physics. The absence of measurements of CP asymmetries in any of the annihilation-dominated modes in particular implies a chance for discovery. Any contribution here is as yet essentially untested by other measurements. The huge ranges allowed in the $b\to d$ modes mean that already very modest measurements in these modes would constrain the global fit further, including the CP asymmetries in the $b\to s$ modes. The latter allow for sizeable BSM contributions, with an already strongly constrained SM background.
\end{enumerate}

These observations show that the framework that has been developed is conservative in its treatment of subleading contributions, but still sufficiently predictive to allow for precision SM determinations as well as searches for BSM physics. The key to improving this analysis are additional measurements of both branching fractions and CP asymmetries in \BDD.

\section{Conclusions \label{sec:conclusions}}

\BDD{} decays provide powerful probes of the SM and physics beyond. Their analysis is complicated by the absence of a reliable calculation from first principles of the corresponding hadronic matrix elements, which do not factorise in the heavy-quark limit \cite{Beneke:1999br,Beneke:2000ry}. We therefore do not attempt to calculate them, but determine them in an \SU-symmetry driven approach from fits to the data, including the relevant symmetry-breaking corrections~\cite{Jung:2014jfa}. The resulting reduced matrix elements are then grouped into topological amplitudes, which allows the identification of additional suppression factors, for instance in an $1/N_C$ expansion. While the individual suppression factors can be sizeable, their combination in terms of a generic factor $\epsilon\sim30\%$ exhibits a hierarchy that is useful in treating the numerous subleading amplitudes. Importantly, this scheme can be tested in the fits themselves, and we find its assumptions to be justified.

Since our strategy relies heavily on the experimental data for the modes under investigation themselves, their treatment is particularly important. We therefore carry out in a first step a detailed analysis of the available experimental data (see Tables~\ref{tab:BRsJD} and \ref{tab:CPasym-exp}), extracting in particular the pertinent branching fractions including their correlations. Several correlations are very large and hence their inclusion is mandatory. From a global fit to these data in our \SU{} framework, we obtain improved and updated predictions for branching fractions and CP asymmetries (c.f. Tables~\ref{tab:BRs-theory} and~\ref{tab:CP-asymmetries} and Figs~\ref{fig:plotBR}, \ref{fig:1D-ACPs}, \ref{fig:1D-SCPs}, \ref{fig:2D-plots}, and \ref{fig:ACP2_ACP7_Combined}). Importantly, we predict $\mathcal B(\bar B^0\to D_s^-D_s^+)$ and several CP asymmetries that are yet to be measured, or have presently much larger uncertainties than our prediction. Those predictions can be tested with already-recorded data, and even more precisely with future data from especially the LHCb and Belle-II experiments.

Our most precise predictions  concern firstly the CP asymmetries in $b\to s$ transitions, which allow for a precision determination of the $B_s$ mixing phase $\phi_s$ through the global fit, and to good approximation through Eq.~(\ref{eq:sinphis}), with a theoretical precision as good as that for the ``golden mode'' $B_s\to J/\psi \phi$, and secondly the quasi-isospin relations Eqs.~(\ref{eq:quasiiso}). The apparent violation of one of the latter has led us to introduce into the fit the possibility of the ratio of production fractions $f_d/f_u$ being different from one. This improves the fit significantly, yielding
 \begin{align}
        (f_d/f_u)^{\text{LHCb,~7 TeV}} &= 0.86\pm0.05\,,
 \end{align} 
corresponding to a significance for a deviation from unity of $2.5\sigma$. Since a confirmation of this value would have significant phenomenological implications, for instance for $B_{(s)}\to \mu^+\mu^-$, additional experimental confirmation is required, both in \BDD{} and in other modes like $B\to J/\psi K$.

In order to probe if a consistent picture emerges at higher precision, measurements of all CP asymmetries of the 10 correlated decay channels are desirable, as well as improved determinations of their branching fractions. The advantage of our approach is that its predictions automatically improve as soon as more data is available.
Especially interesting will be future tests of the quasi-isospin relations Eqs.~(\ref{eq:quasiiso}) and the predicted CP asymmetries. It will also be interesting to analyse potential connections to recent signs of possible anomalies in hadronic decays (see Refs.~\cite{Bordone:2020gao, Iguro:2020ndk, Cai:2021mlt, Bordone:2021cca, Endo:2021ifc,Fleischer:2021cct,Gershon:2021pnc, Lenz:2022pgw, Bhattacharya:2022akr, Amhis:2022hpm, Piscopo:2023opf, Biswas:2023pyw}).
Additional opportunities lie ahead in other $b\to c\bar cD$ ($D=d,s$) transitions like $\bar{B}\rightarrow \bar{D}D^*$ and $\bar{B}\rightarrow \bar{D}^*D^*$ decays, which have many more observables due to their more complex angular dependence (see \emph{e.g.} the recent measurements in Refs.~\cite{LHCb:2019ouq, LHCb:2022uvz, LHCb:2023wbb}), rendering $b\to c\bar cD$ transitions important players in the era of precision flavour physics.

\begin{acknowledgments}
J.D. is supported by the European Research Council under the starting grant  Beauty2Charm 852642.
The work of M.J. is supported in part by the Italian Ministry of University and Research (MUR) under grant PRIN 2022N4W8WR. 
S.S.~is supported by a Stephen Hawking Fellowship from UKRI under reference EP/T01623X/1 and the STFC research grants ST/T001038/1 and ST/X00077X/1.
For the purpose of open access, the authors have applied a Creative Commons Attribution (CC BY) licence to any Authors Accepted Manuscript version arising. This work uses existing data which is available at locations cited in the bibliography.
\end{acknowledgments}

\begin{appendix}

\section{Detailed List of Branching Ratio Input from Experiment \label{sec:appx-inputs}}

Here, expanding on Sec.~\ref{sec:branching-fractions}, we give a detailed list of the experimental branching ratio data inputs that we use in our analysis~\cite{LHCb:2013sad, LHCb:2014scu, LHCb:2023wbb, CLEO:1995psi, BaBar:2006jvx, BaBar:2006uih, Belle:2008doh, Zupanc:2007pu, Belle:2012mef, Belle:2012tsw, CDF:2012xmd}. 
We summarise the branching ratio input in Tables~\ref{tab:br-input-data} and \ref{tab:new_inputs}.

\subsection{LHCb}
We consider three sets of branching ratio measurements by LHCb given in Refs.~\cite{LHCb:2023wbb, LHCb:2013sad, LHCb:2014scu}. 
We define ratios in which inputs like the $D$ branching ratios and production fractions are made explicit, including the
relative production fraction $f_d/f_u$, which is assumed to be unity in Refs.~\cite{LHCb:2013sad, LHCb:2014scu}:
\begin{align}
R_1 &= \frac{f_s}{f_d}\frac{\mathcal{B}(\bar B_s\to D^+D^-)}{\mathcal{B}(\bar B^0\to D^+D^-)}
=\epsilon_{\rm rel} \kappa \frac{N_{\bar B_s\to D^+D^-}}{N_{\bar B^0\to D^+D^-}}\,,\\
R_2 &= \frac{f_s}{f_d}\frac{f_d}{f_u}\frac{1}{\epsilon'_{\rm rel}}\frac{\mathcal{B}(\bar B_s\to D^0\bar D^0)}{\mathcal{B}(B^-\to D^0
D_s^-)}
=\kappa \frac{N_{\bar B_s\to \bar D^0D^0}}{N_{B^-\to \bar D^0D_s^-}},\\
R_3 &= \frac{f_d}{f_u}\frac{1}{\epsilon'_{\rm rel}}\frac{\mathcal{B}(\bar B^0\to \bar D^0 D^0)}{\mathcal{B}(B^-\to D^0
D_s^-)}
=\frac{N_{\bar B^0\to \bar D^0D^0}}{N_{B^-\to \bar D^0D_s^-}}\,,\\
R_4 &= \frac{f_s}{f_d}\frac{\mathcal{B}(D_s^+\to K^+K^-\pi^+)}{\mathcal{B}(D^+\to K^-\pi^+\pi^+)}\frac{\mathcal{B}(\bar B_s\to
D_s^+D_s^-)}{\mathcal{B}(\bar B^0\to D^+D_s^-)}\nonumber\\
&= \epsilon_{\rm rel}^{B_0/B_s}\kappa \frac{N_{\bar B_s\to D_s^-D_s^+}}{N_{\bar B^0\to D^+D_s^-}}\,,\\
R_5 &= \frac{f_u}{f_d}\frac{\mathcal{B}(D^0\to K^-\pi^+)}{\mathcal{B}(D^+\to K^-\pi^+\pi^+)}\frac{\mathcal{B}(B^-\to
D^0D_s^-)}{\mathcal{B}(\bar B^0\to D^+D_s^-)}\nonumber\\
&= \epsilon_{\rm rel}^{B^-/B^0}\kappa \frac{N_{B^-\to D^0D_s^-}}{N_{\bar B^0\to D^+D_s^-}}\,,\label{eq:R5}\\
R_6 &= \frac{f_s}{f_d}\frac{\mathcal{B}(\bar B_s\to D_s^-D^+)}{\mathcal{B}(\bar B^0\to D^+D_s^-)}
=\epsilon_{\rm rel} \frac{N_{\bar B_s\to D_s^-D^+}}{N_{\bar B^0\to D^+D_s^-}}\,,\label{eq::BcB0}\\
R_7 &= \frac{\mathcal{B}(B^-\rightarrow D^-D^0) \mathcal{B}(D^-\rightarrow K^+\pi^-\pi^-)}{\mathcal{B}(B^-\rightarrow D_s^-D^0) \mathcal{B}(D_s^-\rightarrow K^+K^- \pi^-)}\,.
\end{align}
The efficiencies and counting rates are given in Ref.~\cite{LHCb:2013sad}, as well as the systematic uncertainties and their
correlations\footnote{
Note that when reproducing the numerical results for the ratio of branching ratios in Eq.~\eqref{eq:R5} given in Ref.~\cite{LHCb:2013sad} from the number of events given therein, we observe a mismatch of $\sim 2\%$, for which we account with a corresponding correction factor. Note further the presence of misprints in the table for the systematic uncertainties affecting the decay channel $\bar{B}^0\rightarrow \bar{D}^0 D^0$.}. 
The factor $\epsilon'_{\rm rel}$ contains the relevant $D$ branching fractions for these modes. The result for the ratio $R_6$ in Ref.~\cite{LHCb:2013sad} is
superseded by the result in Ref.~\cite{LHCb:2014scu}. Since it is essentially uncorrelated to the other ratios, we simply replace it by the newer result. 

\subsection{CDF}

The relevant measurement from CDF can be found in Ref.~\cite{CDF:2012xmd}. The quantity measured is essentially the ratio $R_{4}$ defined
above, where one of the $D_s$ mesons in the numerator and the one in the denominator are reconstructed in the same way,
while the second $D_s$ in the numerator is reconstructed via $D_s^+\to \phi(\to K^+K^-)\pi^+$. The ratio of production
fractions $f_s/f_d$ has been shown to depend on $p_T$ and hence the experiment, so we treat this quantity as independent
for the Tevatron. We hence define the observable 
\begin{align}
R_{4}^{\rm CDF} &= \left.\frac{f_s}{f_d}\right|_{\rm Tev}\frac{\mathcal{B}(D_s^+\to \phi(\to K^+K^-)\pi^+)}{\mathcal{B}(D^+\to K^-\pi^+\pi^+)} \frac{\mathcal{B}(\bar B_s\to
D_s^+D_s^-)}{\mathcal{B}(\bar B^0\to D^+D_s^-)}\,.
\end{align}
The numerical value in Table~\ref{tab:br-input-data} is calculated from the information given in the paper and information from
the PDG for the corresponding year.

\subsection{CLEO~II}

We use the information from the analysis in Ref.~\cite{CLEO:1995psi}. This analysis makes the dependence on the main $D$ branching fractions
explicit, which we use to update them. There is still a residual dependence on other $D$ decay modes, which we cannot
update since the corresponding dependence is not made explicit and cannot be easily reconstructed. We introduce the quantities
\begin{align}
\mathcal B^{\rm CLEO}_1 &= 2 f_{+-}\mathcal{B}(B^-\to D^0 D_s^-)\mathcal{B}(D^0\to K^-\pi^+) \mathcal{B}(D_s^-\to \phi\pi^-)\,,\\
\mathcal B^{\rm CLEO}_2 &= 2 f_{00}\mathcal{B}(\bar B^0\to D^+ D_s^-)\mathcal{B}(D^+\to K^-\pi^+\pi^+) \mathcal{B}(D_s^-\to \phi\pi^-)\,,
\end{align}
which can again be calculated from the information in the paper.

\subsection{BaBar}

The two relevant analyses by BaBar are given in Refs.~\cite{BaBar:2006jvx, BaBar:2006uih}. The analysis Ref.~\cite{BaBar:2006jvx} is special in that it uses a double-tagging technique such that their results do not depend
on the $\Upsilon$ branching fractions, thereby providing absolute branching fraction measurements. They make explicit
the $D_s^+\to \phi\pi^+$ branching fraction since it constitutes the largest source of systematic uncertainty, while the other $D$ branching fractions cannot be updated by us. We therefore use two of the measurements directly ($\mathcal B^{\rm BaBar, dir}_{1,2}$), while two more
are rescaled by a factor $k=3.6\%/\mathcal{B}(D_{s}^+\to\phi\pi^+)$ ($\mathcal B^{\rm BaBar, re}_{1,2}$):
\begin{align}
\mathcal B^{\rm BaBar, dir}_1 &= \mathcal{B}(B^-\to D^0D_s^-)\,,\\
\mathcal B^{\rm BaBar, dir}_2 &= \mathcal{B}(\bar B^0\to D^+D_s^-)\,,\\
\mathcal B^{\rm BaBar, re}_1 &= \mathcal{B}(B^-\to D^0D_s^-)/k\,,\\
\mathcal B^{\rm BaBar, re}_2 &= \mathcal{B}(\bar B^0\to D^+D_s^-)/k\,.
\end{align}
The $k$ factor does not enter all of the measurements, since only one of the $D$ mesons in each decay is explicitly reconstructed. This factor is only present for cases when the $D_s$ is the meson that is reconstructed and is absent for the cases where the $D^{(0,+)}$ is reconstructed. For Ref.~\cite{BaBar:2006uih}, we define
\begin{align}
\mathcal B^{\rm BaBar}_3 &= 2f_{00} \mathcal{B}(D^+\to K^-\pi^+\pi^+)^2\mathcal(\bar B^0\to D^+D^-)\,,\\
\mathcal B^{\rm BaBar}_4 &= 2f_{+-} \mathcal{B}(D^+\to K^-\pi^+\pi^+)\mathcal{B}(B^-\to D^0D^-)\,,
\end{align}
where only the charged $D$ decay is rescaled, since several $D^0$ decays are used that cannot be updated from the information in the paper. 

\subsection{Belle}

We include four different analyses from Belle, Refs.~\cite{Zupanc:2007pu, Belle:2008doh,  Belle:2012mef, Belle:2012tsw}. For the analysis in Ref.~\cite{Zupanc:2007pu}, we define
\begin{align}
\mathcal B_{2a}^{\rm Belle} &= 2f_{00}\mathcal{B}(\bar B^0\to D^+D_s^-)\mathcal{B}(D_s^-\to\phi(\to K^+K^-)\pi^-)\,,\\
\mathcal B_{2b}^{\rm Belle} &= 2f_{00}\mathcal{B}(\bar B^0\to D^+D_s^-)\mathcal{B}(D_s^-\to K^{*0}(\to K^-\pi^+)K^-)\,,\\
\mathcal B_{2c}^{\rm Belle} &= 2f_{00}\mathcal{B}(\bar B^0\to D^+D_s^-)\mathcal{B}(D_s^-\to K_S(\to \pi^+\pi^-)K^-)\,.
\end{align}
These three measurements have sizeable common systematic uncertainties. We construct a correlation matrix from the
information in the paper. As a crosscheck we confirm that our procedure reproduces the average given in Ref.~\cite{Zupanc:2007pu}. For the analysis in Ref.~\cite{Belle:2008doh},
we do not have the possibility to update the $D$ branching fractions, but here they are not the dominant uncertainties.
We  therefore only include the production fraction $f_{+-}$:
\begin{align}
\mathcal B_1^{\rm Belle} &= 2 f_{+-}\mathcal{B}(B^-\to D^0D^-)\,.
\end{align}

For the analysis in Ref.~\cite{Belle:2012mef}, 
superseding Ref.~\cite{Fratina:2007}, 
we define 
\begin{align}
\mathcal B_3^{\rm Belle} &= 2f_{00}\mathcal{B}(\bar B^0\to D^+D^-)\mathcal{B}(D^+\to K^-\pi^+\pi^+)^2\,,
\end{align}
which constitutes an approximate rescaling, since one of the $D$ mesons is also reconstructed in a second decay mode,
albeit only in $\sim 25\%$ of cases. Finally, Belle measured also an absolute $B_s$ branching fraction~\cite{Belle:2012tsw}
\begin{align}
\mathcal{B}_4^{\mathrm{Belle}} = \mathcal{B}(B^0_s \to D^+_s D^-_s)\,.
\end{align}
The result suffers from a large uncertainty
in the number of $B_s$ mesons, which appears only in this measurement. It is not possible for us to update the $D_s$
branching fractions used, so we use the value as given in Ref.~\cite{Belle:2012tsw}.

\begin{table}
    \centering
    \begin{tabular}{c|c}\hline\hline
       Input  & Value\\
       \hline
       $R_1 = \frac{f_s}{f_d}\frac{\mathcal{B}(\bar B_s\to D^+D^-)}{\mathcal{B}(\bar B^0\to D^-D^+)}$  & $(2.76 \pm 0.52) \times 10^{-1}$ \cite{LHCb:2013sad}\\
       $R_2 = \frac{f_s}{f_d}\frac{f_d}{f_u}\frac{1}{\epsilon'_{\rm rel}}\frac{\mathcal{B}(\bar B_s\to \bar D^0D^0)}{\mathcal{B}(B^-\to D^0
D_s^-)}$  & $(9.24 \pm 1.01) \times 10^{-3}$ \cite{LHCb:2014scu} \\ 
       $R_3 = \frac{f_d}{f_u}\frac{1}{\epsilon'_{\rm rel}}\frac{\mathcal{B}(\bar B^0\to \bar D^0 D^0)}{\mathcal{B}(B^-\to D_s^- D^0)}$ &  $(2.52 \pm 1.17) \times 10^{-3}$ \cite{LHCb:2013sad}\\ 
       $R_4 = \frac{f_s}{f_d}\frac{\mathcal{B}(D_s^+\to K^+K^-\pi^+)}{\mathcal{B}(D^+\to K^-\pi^+\pi^+)}\frac{\mathcal{B}(\bar B_s\to
D_s^-D_s^+)}{\mathcal{B}(\bar B^0\to D_s^-D^+)}$ & $(8.59 \pm 0.01) \times 10^{-2}$ \cite{LHCb:2013sad}\\
       $R_5 = \frac{f_u}{f_d}\frac{\mathcal{B}(D^0\to K^-\pi^+)}{\mathcal{B}(D^+\to K^-\pi^+\pi^+)}\frac{\mathcal{B}(B^-\to
D_s^-D^0)}{\mathcal{B}(\bar B^0\to D^-D_s^+)}$  & $(5.19 \pm 0.29) \times 10^{-1}$ \cite{LHCb:2013sad}\\ 
       $R_6 = \frac{f_s}{f_d}\frac{\mathcal{B}(\bar B_s\to D_s^-D^+)}{\mathcal{B}(\bar B^0\to D_s^-D^+)}$  & $(9.79 \pm 1.74) \times 10^{-3}$ \cite{LHCb:2013sad}\\
       $R_7 = \frac{\mathcal{B}(B^-\rightarrow D^-D^0) \mathcal{B}(D^-\rightarrow K^+\pi^-\pi^-)}{\mathcal{B}(B^-\rightarrow D_s^-D^0) \mathcal{B}(D_s^-\rightarrow K^+K^- \pi^-)}$ & $(7.25\pm 0.13)\times 10^{-2}$ \cite{LHCb:2023wbb}\\
       $\mathcal{B}^{\mathrm{CLEO}}_1 = 2 f_{+-}\mathcal{B}(B^-\to D_s^- D^0)\mathcal{B}(D^0\to K^-\pi^+) \mathcal{B}(D_s^-\to \phi\pi^-)$  & $(1.72 \pm 0.46) \times 10^{-5}$ \cite{CLEO:1995psi}\\ 
       $\mathcal{B}^{\mathrm{CLEO}}_2 = 2 f_{00}\mathcal{B}(\bar B^0\to D_s^- D^+)\mathcal{B}(D^+\to K^-\pi^+\pi^+) \mathcal{B}(D_s^-\to \phi\pi^-)$  & $(2.80 \pm 1.00) \times 10^{-5}$ \cite{CLEO:1995psi} \\ 
       $\mathcal{B}^{\mathrm{BaBar}}_1$ (Dir.) $= \mathcal{B}(B^-\to D_s^-D^0)$ & $(9.0 \pm 2.28) \times 10^{-3}$ \cite{BaBar:2006jvx}\\ 
       $\mathcal{B}^{\mathrm{BaBar}}_1$ (Re.) $= \mathcal{B}(B^-\to D_s^-D^0)/k$ & $(7.4 \pm 2.1) \times 10^{-3}$ \cite{BaBar:2006jvx}\\ 
       $\mathcal{B}^{\mathrm{BaBar}}_2$ (Dir.) $= \mathcal{B}(\bar B^0\to D_s^-D^+)$ & $(1.33 \pm 0.37) \times 10^{-2}$ \cite{BaBar:2006jvx}\\
       $\mathcal{B}^{\mathrm{BaBar}}_2$ (Re.) $= \mathcal{B}(\bar B^0\to D_s^-D^+)/k$  & $(1.11 \pm 0.24) \times 10^{-2}$ \cite{BaBar:2006jvx}\\
       $\mathcal{B}^{\mathrm{BaBar}}_3 = 2f_{00} \mathcal{B}(D^+\to K^-\pi^+\pi^+)^2\mathcal{B}(\bar B^0\to D^-D^+)$  & $(2.37 \pm 0.40) \times 10^{-6}$ \cite{BaBar:2006uih}\\  
       $\mathcal{B}^{\mathrm{BaBar}}_4 = 2f_{+-} \mathcal{B}(D^+\to K^-\pi^+\pi^+)\mathcal{B}(B^-\to D^-D^0)$ & $(3.50 \pm 0.65) \times 10^{-5}$ \cite{BaBar:2006uih}\\ 
       $\mathcal{B}_{1}^{\mathrm{Belle}} = 2 f_{+-}\mathcal{B}(B^-\to D^-D^0)$  & $(3.85 \pm 0.49) \times 10^{-4}$ \cite{Belle:2008doh}\\ 
       $\mathcal{B}_{2a}^{\mathrm{Belle}} = 2f_{00}\mathcal{B}(\bar B^0\to D_s^-D^+)\mathcal{B}(D_s^-\to\phi(\to K^+K^-)\pi^-)$   & $(1.68 \pm 0.21) \times 10^{-4}$ \cite{Zupanc:2007pu}\\ 
       $\mathcal{B}_{2b}^{\mathrm{Belle}} = 2f_{00}\mathcal{B}(\bar B^0\to D_s^-D^+)\mathcal{B}(D_s^-\to K^{*0}(\to K^-\pi^+)K^-)$  & $(1.83 \pm 0.23)\times 10^{-4}$ \cite{Zupanc:2007pu}\\ 
      $\mathcal{B}_{2c}^{\mathrm{Belle}} = 2f_{00}\mathcal{B}(\bar B^0\to D^-D^+)\mathcal{B}(D^+\to K^-\pi^+\pi^+)^2$   & $(7.60 \pm 1.03) \times 10^{-5}$ \cite{Zupanc:2007pu}\\
      $\mathcal{B}_{3}^{\mathrm{Belle}} = 2f_{00}\mathcal{B}(\bar B^0\to D^-D^+)\mathcal{B}(D^+\to K^-\pi^+\pi^+)^2$   & $(1.87 \pm 0.20) \times 10^{-6}$ \cite{Belle:2012mef}\\
      $\mathcal{B}_{4}^{\mathrm{Belle}} = \mathcal{B}(B^0_s \to  D^-_sD^+_s)$  & $(5.90 \pm 1.64) \times 10^{-3}$ \cite{Belle:2012tsw}\\
      $R_{4}^{\mathrm{CDF}} = \left.\frac{f_s}{f_d}\right|_{\rm Tev}\frac{\mathcal{B}(D_s^+\to \phi(\to K^+K^-)\pi^+)}{\mathcal{B}(D^+\to K^-\pi^+\pi^+)} \frac{\mathcal{B}(\bar B_s\to
D_s^-D_s^+)}{\mathcal{B}(\bar B^0\to D_s^-D^+)}$  & $(4.65 \pm 0.61) \times 10^{-2}$ \cite{CDF:2012xmd}\\
       $\frac{f_s}{f_d}\mathcal{B}(D_s^+ \to K^+ K^- \pi^+)$ & $(1.44 \pm 0.10) \times 10^{-2}$ \cite{Bordone:2020gao, LHCb:2011leg, Storaci:2013jqy}\\
       $\epsilon(B^- \to  D_s^-D^0)$ & $(1.66 \pm 0.03) \times 10^{-3}$ \cite{LHCb:2013sad}\\
       $\epsilon(\overline{B}_s^0 \to  \overline{D}^0 (\to K^+ \pi^-)D^0 (\to K^- \pi^+))$ & $(1.90 \pm 0.03) \times 10^{-3}$ \cite{LHCb:2013sad}\\
       $\epsilon(\overline{B}^-_s \to  \overline{D}^0 (\to K^- \pi^+\pi^-\pi^+)D^0 (\to K^- \pi^+))$& $(6.1 \pm 0.2) \times 10^{-4}$ \cite{LHCb:2013sad}\\
       $\mathcal{B}(D_s^- \to K^+ K^- \pi^+)$  & $(5.39 \pm 0.15) \times 10^{-2}$ \cite{PDG2012}\\
       $\mathcal{B}(D^0 \to K^- \pi^+)$ & $(3.95 \pm 0.03) \times 10^{-2}$ \cite{PDG2012}\\
       $\mathcal{B}(D^0 \to K^- \pi^+ \pi^- \pi^+)$ & $(8.22 \pm 0.14) \times 10^{-2}$ \cite{PDG2012}\\
       $\mathcal{B}(D^+ \to K^+ \pi^-\pi^+)$ & $(9.38 \pm 0.16) \times 10^{-2}$ \cite{PDG2012}\\
       $\mathcal{B}(D_s^- \to \phi \pi^-)$  & $(4.5 \pm 0.4) \times 10^{-2}$  \cite{PDG2022}\\
       $\mathcal{B}(D_s^- \to \phi (\to K^+ K^-) \pi^-)$  & $(2.24 \pm 0.08) \times 10^{-2}$ \cite{PDG06}\\
       $\mathcal{B}(D_s^- \to K^{*0} (\to K^+\pi^-) K^-)$  & $(2.58 \pm 0.08) \times 10^{-2}$ \cite{PDG06}\\
       $\mathcal{B}(D_s^- \to K_S^0 (\to \pi^+ \pi^-) K^-)$ & $(1.01 \pm 0.03) \times 10^{-2}$ \cite{CLEO:2006any}\\
       $f_{00}$ & $(4.88 \pm 0.13) \times 10^{-1}$ \cite{Jung:2015yma}\\
       \hline\hline
    \end{tabular}
    \caption{$B$ and $D$ decay branching ratio input data from experiment.}
    \label{tab:br-input-data}
\end{table}

\begin{table}
    \centering
{\renewcommand{\arraystretch}{2}%
\begin{tabular}{ c|c|c } \hline\hline
 Input  & Ref & Correlation\\
\hline
$R_1$  &  \cite{LHCb:2013sad} & \multirow{5}{*}{$\begin{pmatrix} 100 & 2.12 & 0.0 & 4.64 & 0.0 \\ 2.12  & 100 & 0.23  & 4.61  & -1.87  \\ 0.0 & 0.23  & 100 & 0.0 & -0.76 \\ 4.64 & 4.61  & 0.0 & 100 & 3.14 \\
0.0 & -1.87 & -0.76  & 3.14  & 100  \end{pmatrix}$}\\
$R_2$  & \cite{LHCb:2013sad}& \\ 
       $R_3$ &  \cite{LHCb:2013sad}& \\
       $R_4$ &  \cite{LHCb:2013sad}& \\ 
       $R_5$  &  \cite{LHCb:2013sad}& \\
       \hline
       $\mathcal{B}^{\mathrm{CLEO}}_1$  &  \cite{CLEO:1995psi}& \multirow{2}{*}{$\begin{pmatrix} 
       100 & 38\\
      38 & 100 
       \end{pmatrix}$}\\
       $\mathcal{B}^{\mathrm{CLEO}}_2$  &  \cite{CLEO:1995psi}&
       \\ 
       \hline
       $\mathcal{B}_{2a}^{\mathrm{Belle}}$ & \cite{Zupanc:2007pu} & \multirow{3}{*}{$\begin{pmatrix} 
       100 & 81 & 77 \\
       81 & 100 & 77 \\
       77 & 77 & 100
       \end{pmatrix}$}\\
       $\mathcal{B}_{2b}^{\mathrm{Belle}}$  & \cite{Zupanc:2007pu}& \\ 
       $\mathcal{B}_{2c}^{\mathrm{Belle}}$    & \cite{Zupanc:2007pu} & 
       \\ 
\hline\hline
\end{tabular}}
\caption{Correlations in $\%$ for the branching ratio data used.}
\label{tab:new_inputs}
\end{table}

\end{appendix}

\clearpage

\bibliography{B2DD.bib}
\bibliographystyle{apsrev4-1}

\end{document}